\documentclass{article}
\pdfoutput=1
\usepackage[pdftex]{graphicx}
\usepackage{graphicx}
\usepackage{subfigure}
\usepackage{verbatim}
\usepackage{cite}
\usepackage{amssymb}
\usepackage{amsmath}
\usepackage{epstopdf}
\usepackage{float}

\newcommand{\dc}{\mathrm{DC}}
\newcommand{\rf}{\mathrm{RF}}

\newcommand{\urf}{U^{\rf}}
\newcommand{\udc}{U^{\dc}}
\newcommand{\urfij}{U^{\rf}_{ij}}
\newcommand{\udcij}{U^{\dc}_{ij}}

\begin{document}

\title{Stability analysis of surface ion traps}

\author{Arkadas Ozakin\thanks{arkadas.ozakin@gtri.gatech.edu} ~and Fayaz Shaikh\thanks{fayazs@gatech.edu}\\
        \emph{Quantum Information Systems Group}\\
                \emph{Georgia Tech Research Institute}}

\maketitle

\begin{abstract}

Motivated by recent developments in ion trap design and fabrication,
we investigate the stability of ion motion
in asymmetrical, planar versions of the classic Paul trap.
The equations of
motion of an ion in such a trap are generally coupled due to a nonzero
relative angle $\theta$ between the principal axes of RF and DC
fields, invalidating the assumptions behind the standard stability
analysis for symmetric Paul traps. We obtain stability diagrams for
the coupled system for various values of $\theta$,
generalizing the standard $q$-$a$ stability diagrams.
We use multi-scale perturbation theory to obtain approximate
formulas for the boundaries of the primary stability region and obtain
some of the stability boundaries independently by using the method of
infinite determinants. We cross-check the consistency of
the results of these methods.

Our results show that while the primary stability region is quite
robust to changes in $\theta$, a secondary stability region is highly
variable, joining the primary stability region at the special case of
$\theta=45^{\circ}$, which results in a significantly enlarged stability
region for this particular angle.

We conclude that while the stability diagrams for classical, symmetric
Paul traps are not entirely accurate for asymmetric surface traps (or for 
other types of traps with a relative angle between the RF and DC axes), 
they are ``safe''
in the sense that operating conditions deemed stable according to
standard stability plots are in fact stable for asymmetric traps, as well.
By ignoring  the coupling in the equations, one only underestimates
the size of the primary
stability region.

\end{abstract}
\section{Introduction}

\begin{figure}[H]
\centering
\includegraphics[scale=0.5]{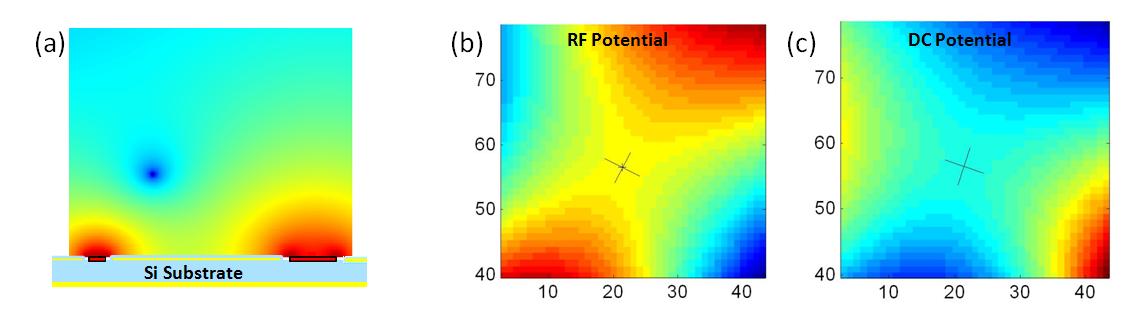}
\caption[]{The cross-section of an example surface trap and
the associated electric fields. Figure (a) shows
the RF electrodes (in red) and the DC control electrodes (in yellow) 
on a substrate and the total pseudopotential. The DC electrodes are used to trap
the ion axially.
The trap axis is orthogonal to the page, and the trap center (the location of the ion)
is at the minimum of the pseudopotential, shown in blue. Figures (b) and (c) show the RF and DC potentials near the trap center,
as indicated. The potentials were calculated
numerically using the trap geometry. As can be seen,
the RF and DC
principal axes are not aligned perfectly, and have a small
angle between them.
Due to this nonzero relative angle, the
ion motion is coupled in the two radial directions, independent
of the orientation of the coordinate axes.
}\label{fig:trap_dcrf_pot}
\end{figure}

In the quest to miniaturize ion traps for quantum information
processing (QIP), many of the recent designs being explored are planar
versions of linear Paul traps.\cite{chiaverini2005surface,kim2007integrated, brady2010integration}
Since the electrodes all lie in a single plane,
these traps can be constructed
using VLSI microfabrication, which offers great
scalability and potential to be integrated with other useful on-chip
components such as mirrors, fiber ferrules, and cavities.
The ions trapped by a surface trap are cooled by laser beams that are
commonly aligned parallel to the trap surface. In order to
cool the ions efficiently, the three principal axes of the
trap pseudopotential must have nonzero components along the laser
direction \cite{wineland1997experimental}. In particular, none of the
principal axes can be vertical, if the cooling beam is
parallel to the horizontal trap surface.  The necessary tilt in the principal
axes relative to the beam direction is typically realized by using asymmetric
electrode designs and/or setting the voltages in an asymmetrical manner.
In general, such
asymmetry, in addition to introducing the desired tilt,
introduces a relative angle between the principal axes of
the RF and DC fields. (See Figure~\ref{fig:trap_dcrf_pot}.)
When the angle between the RF and DC principal
axes is nonzero, the classical equations of motion of an
ion near the trap center are given
by a coupled version of the Mathieu equation.  The
stability properties of such a coupled system cannot be
obtained from the classical stability analysis of Paul traps, which
assumes that the equations are decoupled. Thus, it is not clear, a
priori, that the operating conditions obtained by resorting
to the stability analysis of symmetric Paul traps
will result in stable ion motion in asymmetric surface traps, as well.
One needs to do the stability analysis
for the asymmetric case from scratch, and obtain the corresponding stable
operating conditions.

In this paper, we generalize the standard $q$-$a$ stability diagrams for
symmetric Paul traps to obtain stability diagrams for asymmetric surface
traps, or more generally, for trap designs and operating conditions 
that result in a relative angle between the radial axes of RF and DC fields.
We also obtain approximate formulas for the
boundaries of the primary
stability region, generalizing the formulas for the symmetric, decoupled
case. These results give the stable operating conditions for asymmetric
surface traps, and can serve as a reference for experimentalists trying
to select regions of stability for ion motion, instead of having to
rely on the untested assumption that
the diagrams for symmetric Paul traps are still applicable. 

We approach the problem by both numerical and
analytical techniques.
On the numerical side, for a given set of parameter values
describing the operating conditions of the
system (such as the generalizations of the $q$ and $a$
parameters of the symmetric Paul trap), we obtain a basis set of
numerical solutions, and by using results from Floquet
theory, decide whether the system is stable or unstable under
the given conditions.
By scanning
a range of values of the parameters and collecting stability data over a region in the parameter space,
we obtain the
relevant stability diagrams. Our analysis confirms the stability theory
prediction~\cite{arnold1989mathematical,seyranian2005coupling}
that as the parameters are varied,
instabilities develop when the eigenvalues of a certain
solution matrix
collide on the unit circle. As an independent check, we utilize the
method of infinite determinants~\cite{hansen1985stability}
to directly obtain some boundaries of the stable regions,
and show that these curves
agree with the results from Floquet theory.

On the analytic side, we utilize \textit{multi-scale perturbation
theory} to obtain approximations to the curves bounding the
primary stability region \cite{nayfeh1979nonlinear,nayfeh1973perturbation}.
These approximate results take the form
of formulas that relate the parameters describing the system, such as
the generalized $q$ and $a$ parameters mentioned above,
and the angle between the RF and DC principal axes.
Certain novelties of the coupled multi-variable case
complicate the analysis,
and
we use a hint from our numerical results to pick the relevant curves.

We show that our analytical results well approximate the boundaries
of the stability domains in their regimes of applicability,
except for the special case of a 45 degree tilt between the RF and DC
principal axes. We comment on this special case, for which the
fundamental stability domain is significantly enlarged.

The paper is organized as follows. In Section~\ref{sec-eom},
we set up the general equations of
motion of the coupled system, restricting our attention to the two-variable case.
In Section~\ref{section:numerical},
we give
a discussion of the stability of periodic Hamiltonian systems, and
then apply the formalism to the specific case of the
coupled Mathieu system, obtaining stability diagrams via numerical
solutions of the equations. In Section~\ref{inf-sec}, we discuss an
alternative method, the method of infinite determinants, which is
capable of obtaining some of the stability boundaries directly,
and demonstrate its
consistency with the results of Section~\ref{section:numerical}.
In Section~\ref{section:multiple}, we turn to the method of multiple
scales, and after a general discussion, apply this method to the
coupled Mathieu system relevant for surface traps. This analysis
results
in
approximate formulas for the boundaries of the primary stability
region, which we check against the numerical results of
Section~\ref{section:numerical}. Finally, in
Section~\ref{section:conclusion}, we summarize our findings, and
discuss the practical implications for ion trap design and operation.

\section{Equations of motion.}\label{sec-eom}

Let us begin by writing the general equations of motion of an ion near the center
of an asymmetrical trap.
We will assume that the oscillating (RF) and static (DC) electric
fields in the ion trap have a coincident zero at the trap center,
which we take to be the origin of our coordinate sysetm,
$\mathbf{x} = (x,y,z) = (0,0,0)$.
We will work in the harmonic approximation and treat
potentials as second order in the displacements from the origin,
and forces (or electric fields) as first order.
\footnote{We work in
the quasi-static approximation, where the RF field is described by an
instantaneous electrostatic potential.}
Since the electric fields vanish at the origin, the first order
terms in the expansions of the potentials vanish. Denoting the potential energy
of an ion by $U = eV$,
where $e$ is the ion charge, and choosing
the zero of $U$ so that $U(\mathbf{0})=0$, we have, up to second order in the
displacements,
\begin{eqnarray}\label{totpot}
  U(x,y,z,t) &=& \urf(x,y,z)\cos{\omega t}+\udc(x,y,z)\\\label{totpot2}
  &=&\frac{1}{2}\sum_{ij}x_i x_j \urf_{ij}\cos(\omega t) + \frac{1}{2}\sum_{ij}
  x_i x_j \udc_{ij}\,,
\end{eqnarray}
where  $x_1, x_2, x_3$ stand for $x, y, z$, respectively, $\omega$ is
the RF angular frequency, and the
$\urf_{ij}$ and $\udc_{ij}$ are the (symmetric) matrices of second derivatives
of the RF and DC potential energies.

The equations of motion are given as,
\begin{eqnarray}
  m\frac{d^2x_i}{dt^2} &=& -\frac{\partial U}{\partial x_i}\\
  &=& -\sum_j \urf_{ij} x_j \cos(\omega t) -\sum_j \udc_{ij}x_j\,.
\end{eqnarray}
Defining a new time variable $\tau$ by $\omega t = 2\tau$ and
denoting the derivatives with respect to $\tau$ by dots, we
get,
\begin{equation}\label{eomgeneral}
  \ddot{x}_i + \sum_j A_{ij} x_j + 2\sum_j Q_{ij} x_j \cos 2\tau = 0\,,
\end{equation}
where,
\begin{equation}\label{AQ-def}
  A_{ij} = {4 \udc_{ij}}/{m\omega^2}\,, \qquad{}
  Q_{ij} = {2 \urf_{ij}}/{m\omega^2}\,,
\end{equation}
are the multi-variable versions of stability parameters $a$ and
$q$ of the Mathieu equation \cite{house2008analytic}.
\textit{stiffness matrix}.
$Q_{ij}$ are traceless,
\begin{equation}
  \sum_i A_{ii} = 0\,, \qquad{} \sum_i Q_{ii} = 0\,.
\end{equation}
\paragraph{The decoupled case.}
One can diagonalize $\udcij$ by an orthogonal transformation $\mathbf{S}$,
\begin{equation}\label{stfm}
  x_i = \sum_j S_{ij}\tilde{x}_j,
\end{equation}
and obtain,
\begin{equation}
  \udc(\tilde{x},\tilde{y},\tilde{z}) = \frac{1}{2} \left(\tilde{U}^{\dc}_{11}\tilde{x}_1^2 +
  \tilde{U}^{\dc}_{22}\tilde{x}_2^2 +
  \tilde{U}^{\dc}_{33}\tilde{x}_3^2 \right)\,,
\end{equation}
where the diagonal matrix $\tilde{U}_{ij}$ is given by
$\tilde{U}_{ij} = \sum_{kl} S_{ik}S_{jl}U_{kl}$. If the RF and DC
principal axes coincide, the same transformation (\ref{stfm}) also
diagonalizes $\urfij$, and we get (dropping the tildes, in order to
simplify the notation),
\begin{eqnarray*}
  U =&   \frac{1}{2} \left(\urf_{11}{x}_1^2 +
  \urf_{22}{x}_2^2 +
  \urf_{33}{x}_3^2 \right)\cos(\omega t)+ \\
  & \frac{1}{2} \left(\udc_{11}{x}_1^2 +
  \udc_{22}{x}_2^2 +
  \udc_{33}{x}_3^2 \right)\,.
\end{eqnarray*}
This is the classical case of a symmetric Paul trap. The equations of motion
resulting from this potential are decoupled; Equations
(\ref{eomgeneral}) and (\ref{AQ-def}) become,
\begin{equation}\label{singlemathieu}
  \ddot{x}_i + (a_i + 2q_i\cos 2\tau)x_i = 0\,,
\end{equation}
where,
\begin{equation}
  a_i = {4 \udc_{ii}}/{m\omega^2},\qquad{} q_i={2
\urf_{ii}}/{m\omega^2}\,.
\end{equation}
Equation (\ref{singlemathieu}) is known as the single-variable
Mathieu equation, and by using the results of the
classical Mathieu stability analysis on each
component separately, one can obtain the regions of joint
stability. This gives the standard $a$-$q$ stability plots for the Paul
trap \cite{major2005charged,ghosh1995ion}.

\paragraph{Reduction to two dimensions.}
In most surface trap designs, the RF electrodes are long, and the
total RF field has non-vanishing
components only in the radial directions, which we denote by $x$
and $y$. If, in addition, the electrodes and voltages are symmetric around the
$z=0$ plane, which is commonly the case, the axial ($z$) motion of the ion is
decoupled from the $x$ and $y$ motions, and is simple harmonic (since
the axial force in this case is given by
a DC field linear in the displacement $z$).
The $x$-$y$ equations of
motion are still of the form (\ref{eomgeneral}),
\begin{equation}\label{eomgeneralcopy}
  \ddot{x}_i + \sum_j A_{ij} x_j + 2\sum_j Q_{ij} x_j \cos 2\tau = 0\,,
\end{equation}
but the matrices $\mathbf{A}$ and $\mathbf{Q}$ are now $2\times 2$.
When
the DC field in the $z$ direction is confining, Gauss's law implies
that the DC field has an anti-confining radial component. In other
words, the trace of $\udc_{ij}$, as restricted to the $x$-$y$
plane, must be negative. Due to (\ref{AQ-def}) the same is true
for the matrix
$\mathbf{A}$. Similarly, since the $z$-component of the RF field is
assumed to be zero identically, Gauss's law enforces the $2\times 2$
version of the matrix $\urf_{ij}$ to be traceless. Hence,
$\mathbf{Q}$ is also traceless.

\paragraph{A convenient coordinate system and parametrization.}
Given the general equations (\ref{eomgeneralcopy}),
one can investigate the stability
properties of the system in terms of the entries of the
(now $2\times 2$) matrices $\mathbf{Q}$ and $\mathbf{A}$.
However, in order to obtain stability
plots that reduce, in the decoupled limit, to the familiar $q$-$a$
stability plots of symmetric Paul traps, we will fix the relative
magnitudes of the entries of $\mathbf{Q}$ and $\mathbf{A}$, and vary
their overall scales. In terms of the actual operating conditions of a
trap, this amounts to fixing the trap geometry and the ratios of the
DC electrode voltages, and changing the overall scale of the voltages on
DC and RF electrodes.

The matrices $\mathbf{A}$ and
$\mathbf{Q}$ are symmetric since they are related to
the symmetric matrices $\urf_{ij}$ and $\udc_{ij}$ of (\ref{totpot})-(\ref{totpot2})
through (\ref{AQ-def}). Thus, we can diagonalize
at least one of $\mathbf{A}$ or $\mathbf{Q}$ by a suitable choice of
our coordinate axes $x$ and $y$.  Let us assume that $\mathbf{A}$
is diagonalized, and write it in the form,
\begin{equation}\label{a-diag}
  \mathbf{A} = a \left(\begin{array}{cc}
    1 & 0\\
    0 & -\alpha
  \end{array}\right)\,,
\end{equation}
where $a$ and $\alpha$ are constants to be determined by the electrode
geometry and the DC voltages. As argued above, $\mathbf{A}$ must have
negative trace due to Gauss's law and the fact that the ion is
confined along the $z$-axis. Thus, we must either have $a> 0$ and
$\alpha > 1$,
or $a<0$ and $\alpha < 1$.\footnote{Having $\alpha <0$ and $a<0$
would correspond to the DC potential
being anti-confining along both radial axes. Although
this is possible, it is not frequently
encountered in the
designs used in practice.}

Recall that $\urf_{ij}$ is symmetric and
traceless.  If we were to use a coordinate system
$(\tilde{x},\tilde{y})$ in which $\urf_{ij}$ is diagonal, the
amplitude of the RF potential energy would have the form,
\begin{equation}\label{diagrfpot}
  \urf(\tilde{x},\tilde{y}) =
  \frac{1}{2}\urf_0 (\tilde{x}^2-\tilde{y}^2)\,,
\end{equation}
and $Q_{ij}$, being related to $\urfij$ through (\ref{AQ-def})
would be given as,
\begin{equation}\label{q-diag}
  \mathbf{Q} = q \left(\begin{array}{cc}
    1 & 0\\
    0 & -1
  \end{array}\right)\,,
\end{equation}
where $q$ is a free parameter. The two coordinate systems,
$(\tilde{x},\tilde{y})$
(in which $\mathbf{Q}$ is diagonal) and $(x,y)$
(in which $\mathbf{A}$ is diagonal) are related by a rotation,
\begin{eqnarray}
  \tilde{x} &=& x \cos{\theta} + y\sin{\theta}\\
  \tilde{y} &=& -x\sin{\theta} + y\cos{\theta}\,,
\end{eqnarray}
where $\theta$ is the angle of rotation.
Substituting these in (\ref{diagrfpot}), we get the RF potential
energy in terms of the coordinates along the DC principal axes. This
gives,
\begin{equation}\label{nondiagrfpot}
  \urf(x,y) =\frac{1}{2}\urf_0 (
  x^2\cos{2\theta} - y^2\cos{2\theta} +2xy\sin{2\theta}
  )\cos 2\tau \,.
\end{equation}
Using (\ref{totpot2}) and (\ref{AQ-def}), we get the resulting $\mathbf{Q}$ matrix as,
\begin{equation}\label{q-nondiag}
  \mathbf{Q} = q \left(\begin{array}{cc}
    \cos{2\theta} & \sin{2\theta}\\
    \sin{2\theta} & -\cos{2\theta}
  \end{array}\right)\,.
\end{equation}
Finally, using (\ref{a-diag}) and (\ref{q-nondiag}), the equations of
motion (\ref{eomgeneralcopy})
become,
\begin{eqnarray}\label{eomfin1}
  \ddot{x} + ax + 2q(cx+sy)\cos{2\tau} &=& 0\\
  \ddot{y} - \alpha ay + 2q(sx-cy)\cos{2\tau} &=& 0\,,\label{eomfin2}
\end{eqnarray}
where for brevity we replaced $\cos{2\theta}$ and $\sin{2\theta}$ by
$c$ and $s$, respectively. The classic case of symmetric Paul traps
where $\theta=0$ is recovered by setting $s=0$ and $c=1$.
Below, we will get $q$-$a$ stability plots for various values of
$\alpha$ and $\theta$.
As mentioned above, fixing $\alpha$ and $\theta$ and varying $q$ and $a$
correspond to fixing the trap geometry and the ratios of the DC
voltages, and changing the overall scale of the RF ($q$) and DC ($a$)
voltages.

We next turn to the investigation of the stability properties of
the coupled equations (\ref{eomfin1})-(\ref{eomfin2}).

\section{Stability of the coupled Mathieu system}\label{section:numerical}
\subsection{Periodic systems and stability}
We begin by reviewing some general aspects of linear, periodic systems
and their stability. The discussion is perhaps a bit abstract, but the bottom
line is the following: The stability properties of the system (\ref{eomgeneralcopy})
is determined by first obtaining a basis set of solutions over one period of the RF
field, and then inspecting the eigenvalues of the matrix formed by joining these
fundamental solutions.

\paragraph{Equivalent first order system.}
Let us first rewrite  (\ref{eomgeneralcopy}) as an equivalent,
first order system by
defining the velocity components as new variables. Letting,
\begin{equation}
  \mathbf{u} = \left[\begin{array}{c}\mathbf{x}\\
      \dot{\mathbf{x}}\end{array}\right]\,,
\end{equation}
where $\mathbf{x}=(x,y)$
is the radial position vector, we can rewrite the radial equations of motion
(\ref{eomgeneralcopy}) as,
\begin{equation}\label{first-order}
  \dot{\mathbf{u}} = \mathbf{G}(\tau)\mathbf{u}\,,
\end{equation}
where,
\begin{equation}\label{specific-g}
  \mathbf{G}(\tau) = \left(\begin{array}{cc}
    \mathbf{0} & \mathbf{I}\\
    -2\mathbf{Q}\cos{2\tau} - \mathbf{A}& \mathbf{0}
\end{array}\right)\,,
\end{equation}
$\mathbf{I}$ denoting the
$2\times 2$ identity matrix. The matrix $\mathbf{G}$ is periodic in time with period $T=\pi$.

Although
our primary interest will be in the two-dimensional case

for which $\mathbf{u}$ is a
4-dimensional vector, much of what we will say below will be valid for a
general Hamiltonian system (\ref{first-order}) with a
periodic $\mathbf{G}(\tau)$.

\paragraph{Fundamental solution matrix.}
In order to explore the long-time stability properties of
(\ref{first-order}), we first obtain a set of fundamental
solutions $\mathbf{u}_i(\tau)$ that form a basis for the space of
all solutions. We take the initial value $\mathbf{u}_i(0)$ of the
$i$th fundamental solution to be the
$i$th column of the $4\times 4$-dimensional identity matrix. In
other words, we take the $i$th component of $\mathbf{u}_i(0)$ to be
$1$, all the other components to be $0$.  Since the system
(\ref{first-order}) is linear, a solution with arbitrary initial
condition $\mathbf{u}(0) = \mathbf{u}_0$ can be obtained as an
appropriate linear combination of these fundamental solutions,
\begin{equation}\label{general-solution}
  \mathbf{u}(\tau) = \sum_i u_{0i} \mathbf{u}_i\,.
\end{equation}
We combine the column vectors $\mathbf{u}_i(\tau)$ into a $4\times 4$ ``fundamental
solution matrix'',

\begin{equation}
  \mathbf{U}(\tau) = [\mathbf{u}_1(\tau)\,\mathbf{u}_2(\tau)\,\mathbf{u}_3(\tau)\,
    \mathbf{u}_{4}(\tau)]\,.
\end{equation}
Since each column of the $\mathbf{U}$ satisfies (\ref{first-order}), $\mathbf{U}$
itself satisfies,
\begin{equation}
  \dot{\mathbf{U}} = \mathbf{G}(\tau)\mathbf{U}\,,
\end{equation}
with the initial condition,
\begin{equation}
  \mathbf{U}(0) = I\,.
\end{equation}
The general solution (\ref{general-solution}) can be written as,
\begin{equation}\label{time-trans-zero}
  \mathbf{u}(\tau) = \mathbf{U}(\tau)\mathbf{u}_0\,.
\end{equation}
If the system (\ref{first-order}) were autonomous, that is, if
$\mathbf{G}(\tau)$ were independent of time, it would be possible to
treat the fundamental solution matrix $\mathbf{U}(\tau)$ as
a ``time translation operator''. An analogue of (\ref{time-trans-zero})
would be valid for an arbitrary initial time $\tau_0$---given initial
conditions $\mathbf{u}(\tau_0) = \mathbf{u}_0$,
one could obtain the solution at time
$\tau_0+\tau$ by applying the map $\mathbf{U}(\tau)$.
However, since the system under consideration is not
autonomous, this is not the case. The matrix $\mathbf{U}$
only gives
time translations from the moment $\tau=0$, as in (\ref{time-trans-zero}).

\paragraph{Mapping at a period.}
Although the system (\ref{first-order}) is not invariant under arbitrary time
translations, it is invariant under translations by one period $T=\pi$ of the
periodic matrix $\mathbf{G}(\tau)$.
Using this fact, it is possible to prove that,

\begin{equation}
    \mathbf{U}(mT) =  \left[\mathbf{U}(T)\right]^m\,.
\end{equation}
Thus, repeated applications of the matrix $\mathbf{U}(T)$,
called the matrix for the ``mapping
at a period'', give relevant information about the long time
behavior of the system (\ref{first-order}). In particular, it can be
shown that the equilibrium solution of (\ref{first-order}),
$\mathbf{u}(\tau)=\mathbf{0}$, is stable if and only if the zero vector $\mathbf{u}_0=\mathbf{0}$,
is stable
under successive applications of $\mathbf{U}(T)$ \cite{arnold1989mathematical,seyranian2005coupling}.\footnote{
In the sense that small deviations from the vector $\mathbf{0}$
remain small under successive applications of $\mathbf{U}(T)$.}
If there is a vector $\mathbf{v}$ such that
$\left[\mathbf{U}(T)\right]^m\mathbf{v}$ for $m=1,2,\ldots$ is an
unbounded sequence of vectors, then there will be an
initial condition of the system
(\ref{first-order}) arbitrarily close to origin,
for which the solution will grow unboundedly.

This connection between the stability of (\ref{first-order})
and the stability under successive applications of
the matrix $\mathbf{U}(T)$
opens the door to testing for the stability of
(\ref{first-order}) by investigating the eigenvalue-eigenvector\footnote{or
more generally, Jordan}
decomposition of the matrix $\mathbf{U}(T)$ \cite{arnold1989mathematical,seyranian2005coupling}.
In particular, the equilibrium is unstable if $\mathbf{U}(T)$ has an
eigenvalue $\lambda$ with $|\lambda|>1$. If the eigenvalues of
$\mathbf{G}(T)$ are not repeated, and if all
have magnitudes less than or
equal to $1$, then the equilibrium is stable.

\paragraph{Eigenvalue spectrum of the mapping at a period for Hamiltonian systems.}
While the results mentioned above are valid for any periodic $\mathbf{G}(\tau)$,
the specific case of (\ref{specific-g})
is obtained from a Hamiltonian system, and the
theory of classical mechanics imposes certain conditions on the
eigenvalues of the relevant $\mathbf{U}(T)$.  Time evolution in
Hamiltonian systems is a canonical transformation, and this
forces $\mathbf{U}(T)$ to satisfy~\cite{arnold1989mathematical},
\begin{equation}\label{canonical}
  \mathbf{U}^T(T) \mathbf{J} \mathbf{U}(T) = \mathbf{J},
\end{equation}
where $\mathbf{J}$
is an antisymmetric matrix
given in terms of the $2\times 2$ identity matrix $\mathbf{I}$ as,
\begin{equation}
  \mathbf{J} = \left[\begin{array}{cc}
      \mathbf{0} & -\mathbf{I}\\
      \mathbf{I} & \mathbf{0}
    \end{array}\right].
\end{equation}
Using (\ref{canonical}), it is possible to show that if $\lambda$ is
an eigenvalue of $\mathbf{U}(T)$, then, so is
$\lambda^{-1}$. Since the coefficients of the characteristic
polynomial of $\mathbf{U}(T)$ are all real, we see that $\bar{\lambda}$,
the complex conjugate of $\lambda$, is also an eigenvalue. Thus,
the eigenvalues of $\mathbf{U}(T)$ come in four kinds of
groups:

\begin{itemize}
  \item 4-tuples $\lambda$, $\bar{\lambda}$, $1/\lambda$,
    $1/\bar{\lambda}$, where $\lambda\ne\bar{\lambda}$,
    $|\lambda|\ne1$.
  \item Real pairs, $\lambda$, $1/{\lambda}$, where
    $\lambda=\bar{\lambda}$, $|\lambda|\ne 1$.
  \item Pairs on the unit circle, $\lambda$,
  $\bar{\lambda}$, where $1/{\lambda}=\bar{\lambda}$
  \item Lone eigenvalues, $\lambda =1$.
\end{itemize}
The multiplicities of a given 4-tuple (or a given pair in the real or
unit length cases) are the same.

It follows from this list of possibilities that the trivial solution
$\mathbf{u}=\mathbf{0}$ is stable under successive applications of $\mathbf{U}(T)$
only if all the eigenvalues of $\mathbf{U}(T)$ are on the unit
circle; the remanining cases must have at least one eigenvalue with
magnitude larger than $1$. If, in addition to being of unit magnitude,
all the
eigenvalues are distinct, then the system is necessarily
stable.\footnote{If some of the eigenvalues are degenerate, one has to
consider the Jordan decomposition
\cite{seyranian2005coupling, arnold1989mathematical}.}

As we change the operating conditions of the trap, the
parameters describing the system (\ref{first-order}), i.e.,
the entries of the matrices $\mathbf{Q}$ and
$\mathbf{A}$ change. This results in a change in the solutions, and hence,
the eigenvalues of the mapping at a period,
$\mathbf{U}(T)$. Suppose we start
with a set of distinct eigenvalues on the unit circle, corresponding to a
stable operating condition of the trap.
The above list of possibilities ensures
that as we tweak the parameters (the voltages),
the eigenvalues of $\mathbf{U}(T)$ can move off the unit circle
only if they ``collide''
on the unit circle first, meaning that instabilities develop
through collisions of eigenvalues of unit magnitude.
We will verify this prediction below, for the case of the coupled
Mathieu equations.\footnote{Note that not all such collisions
of eigenvalues result
in instabilities. A more detailed discussion of these issues is given
in \cite{arnold1989mathematical} and \cite{seyranian2005coupling}.}

\subsection{Application to the Mathieu system}\label{numstab}

We next apply the general results reviewed above to
the coupled Mathieu system, (\ref{eomfin1})-(\ref{eomfin2}).
We proceed as follows. For given, specific values for
$\alpha$ and $\theta$, we loop over pre-chosen ranges of $q$ and $a$.
For each pair of values of $q$ and $a$,
we solve the equations
(\ref{eomfin1})-(\ref{eomfin2}) numerically, for four different initial
conditions; we set one of $x(0)$, $y(0)$, $\dot{x}(0)$, $\dot{y}(0)$
to one,
and all
the others to zero. After obtaining the four solutions over one
period of the oscillating RF field (i.e., in terms of the time parameter
$\tau$, over a time period of $T = \pi$), we obtain
$\mathbf{U}(T)$,
the
``mapping at a period'', by joining the four column vectors $\mathbf{u}_i$,
\begin{equation}
   \mathbf{u}_i(T) =
    \left[\begin{array}{c}x_i(T)\\y_i(T)\\\dot{x}_i(T)\\\dot{y}_i(T)\end{array}\right]\,.
\end{equation}
We then
obtain the eigenvalues of this square matrix, and determine the stability properties
of the system by checking the magnitudes of the eigenvalues. According to the discussion above,
if all eigenvalues are distinct and are of unit magnitude, then the system is stable,
and if there is an eigenvalue of magnitude larger than one, the system is unstable.
We go beyond distinguishing between instability and stability, and also note whether
the system is partially stable, having one pair of eigenvalues on the unit circle, and
one pair away from it. While partial instability means instability in real life (since
exciting only the stable subspace of solutions is impossible due to noise), the distinction
between partial instability and full instability will help us below in our discussion of
the multi-scale perturbation analysis.

Looping over values of $q$ and $a$ and recording the stability properties of the system for
each pair of values, we obtain
a stability diagram for given values of $\alpha$ and $\theta$.
We present two such plots below in
Figures \ref{fig:allmultip-0} and \ref{fig:allmultip-6-4}, the first
one being an example of the classical, decoupled Mathieu system,
and the second
one being an example of the coupled case. As mentioned above, we
go beyond the binary classification of stable vs. unstable points, and
indicate regions of partial stability by using a shade of gray.

In Figures
\ref{fig:transitions-0-1}-\ref{fig:transitions-0-2}
and
\ref{fig:transitions-6-4-1}-\ref{fig:transitions-6-4-2},
we
show how the eigenvalues evolve as one moves on the stability plot,
moving across boundaries of stability. As predicted in our discussion
above, changes in the degree of stability are accompanied by
collisions of the eigenvalues on the unit circle.
We will give a more comprehensive set of
stability plots after we present an approximate analytical method (the
method of multiple scales) for obtaining the stability boundaries.

\begin{figure}[H]
\centering
\includegraphics[scale=0.6]{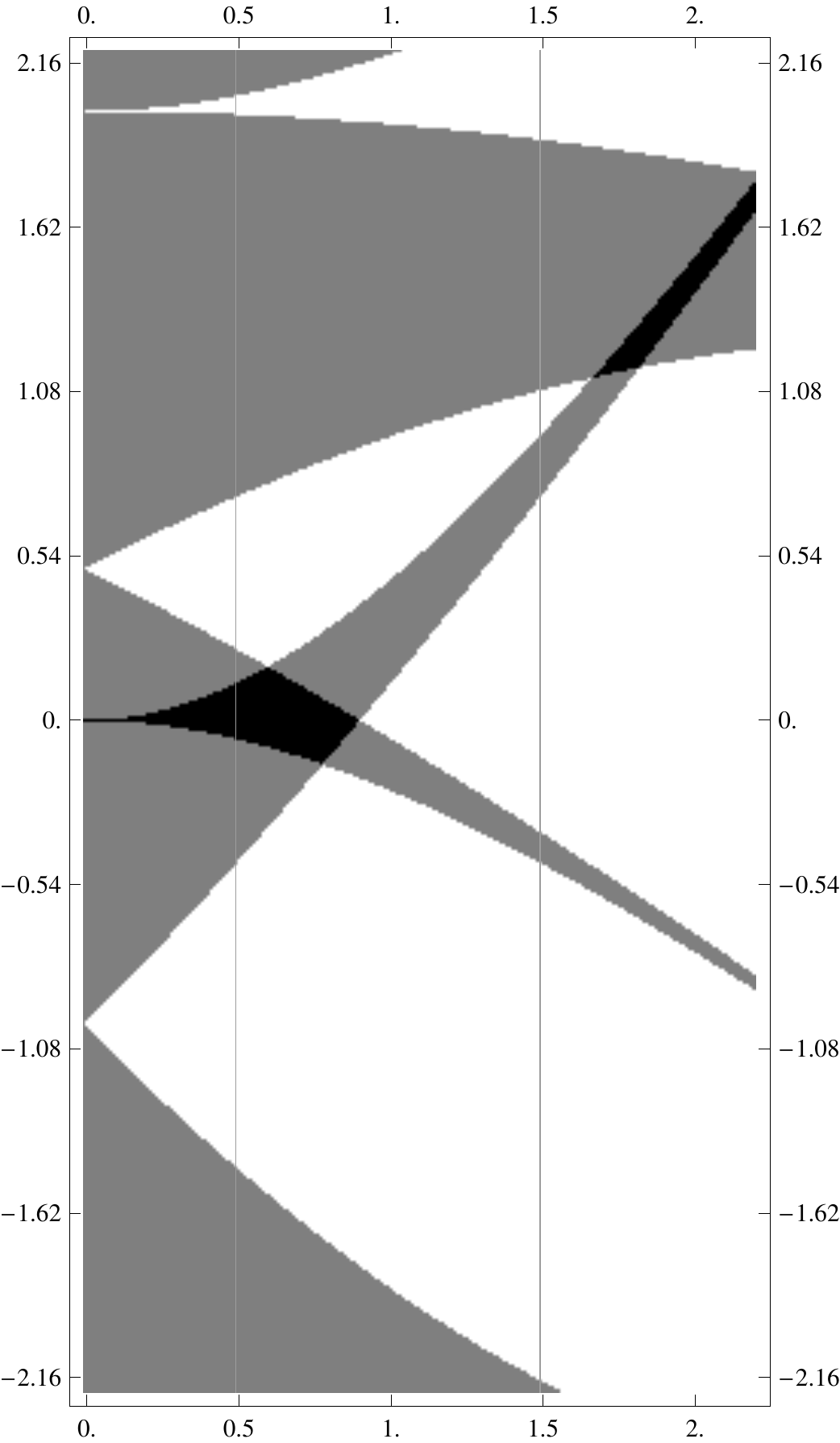}
\caption[]{Stability plot for the decoupled system; $\theta=0^{\circ}$,
  $\alpha = 1/2$. The stability for each value of $q$ ($x$-axis)
  and $a$ ($y$-axis) is
  indicated by the gray level, which represents the
  number of unit length eigenvalues of $\mathbf{U}(T)$, the ``mapping at a
  period''.
  Black indicates complete stability, with all four eigenvalues
  being on the unit circle. Gray indicates partial stability, with
  2 eigenvalues on the unit circle, two away from it. White
  indicates full instability, with all four eigenvalues being off the
  unit circle. The two vertical lines will be referred to in Figures
  \ref{fig:transitions-0-1}
  and \ref{fig:transitions-0-2}}\label{fig:allmultip-0}
\end{figure}

\begin{figure}[H]
\centering
\includegraphics[scale=0.4]{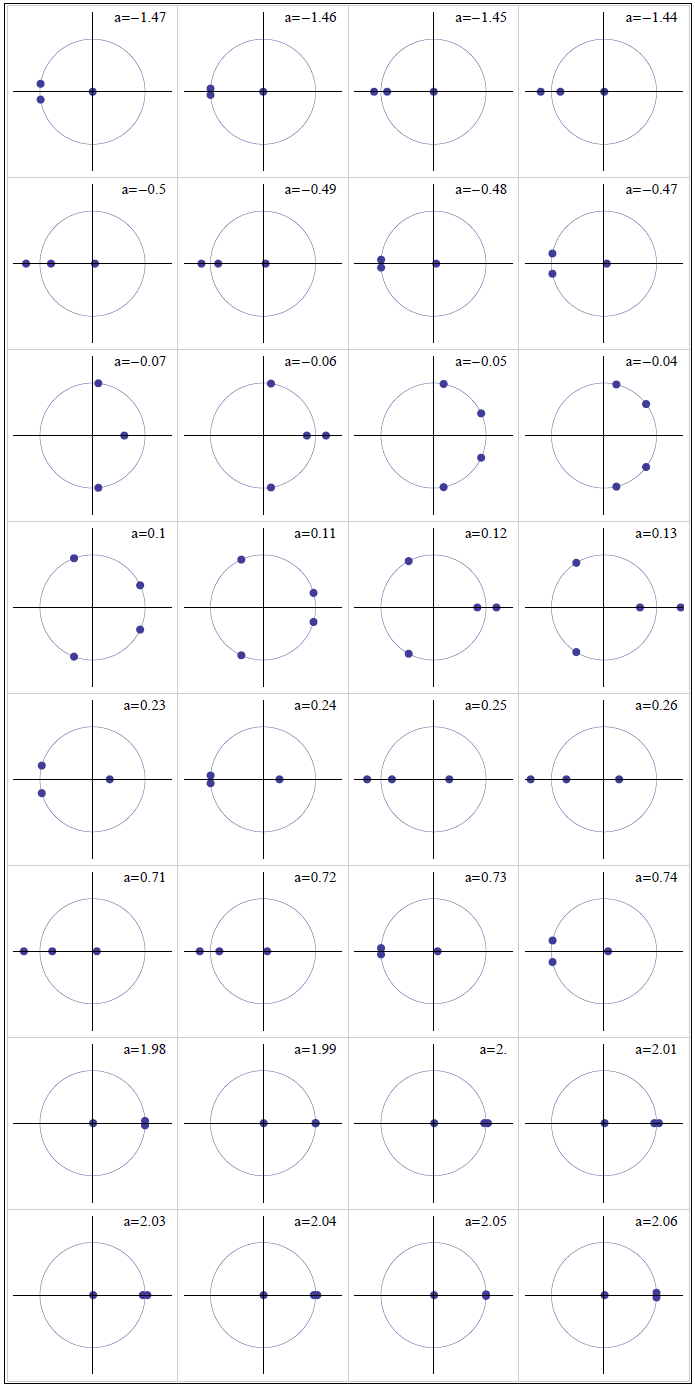}
\caption[]{The evolution of the eigenvalues of the ``mapping at a
  period'', as one moves along the first vertical line (near $q=0.5$) in
  Figure~\ref{fig:allmultip-0}. Each row of four plots corresponds to
  a change in the number of stable eigenvalues as one moves along the first
  vertical line in Figure \ref{fig:allmultip-0}.
  As described in the text, a change in
  the number of stable eigenvectors is effected by a ``collision'' of
  eigenvalues on the unit circle, with eigenvalues on the unit circle
  leaving, or eigenvalues off the unit circle getting on the unit
  circle. This behavior is confirmed in these figures.
  In some plots, one of the four eigenvalues is outside the
  region shown, so only three eigenvalues are seen. Note that in this decoupled
  case, all collisions happen on the real line.}\label{fig:transitions-0-1}
\end{figure}

\begin{figure}[H]
\centering
\includegraphics[scale=0.5]{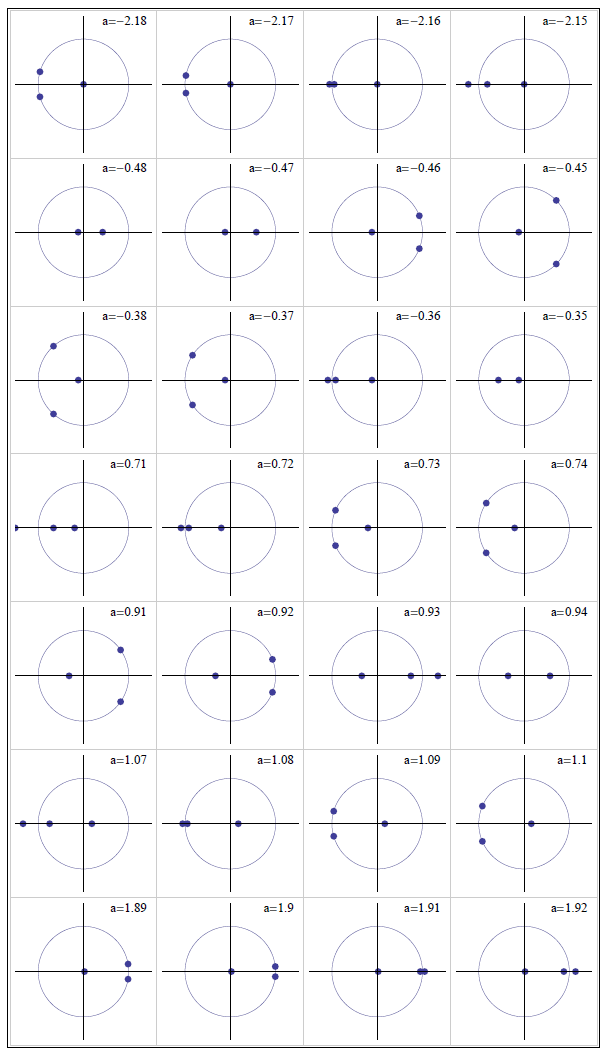}
\caption[]{Similar to Figure~\ref{fig:transitions-0-1},
the evolution of the eigenvalues of the ``mapping at a
  period'', as one moves along the \textit{second} vertical line (near $q=1.5$) in
  Figure~\ref{fig:allmultip-0}. Once again, for this decoupled case,
  collisions happen only on the real line.}\label{fig:transitions-0-2}
\end{figure}

\begin{figure}[H]
\centering
\includegraphics[scale=0.6]{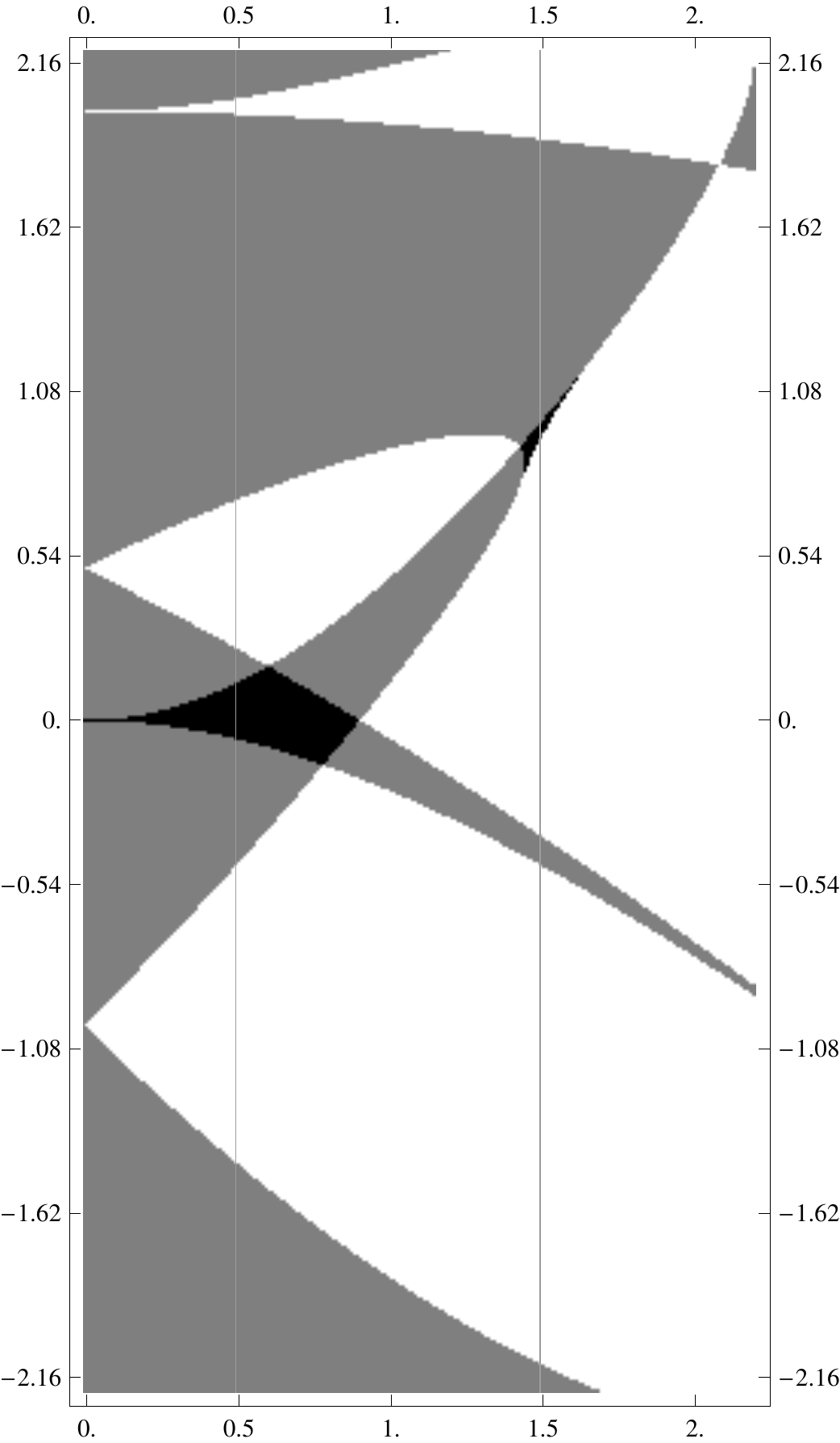}
\caption[]{Stability plot for $\theta=6.4^{\circ}$,
$\alpha = 1/2$. Due to the nonzero angle between RF and DC principal
  axes, the $x$ and $y$ motions are now coupled, and this case
cannot be investigated by the standard, single-variable Mathieu
techniques.
As in Figure~\ref{fig:allmultip-0},
the number of stable
  eigenvalues of $\mathbf{U}(T)$ is
  indicated by the gray level, black corresponding to complete
stability and white corresponding to complete instability.}\label{fig:allmultip-6-4}
\end{figure}

\begin{figure}[H]
\centering
\includegraphics[scale=0.4]{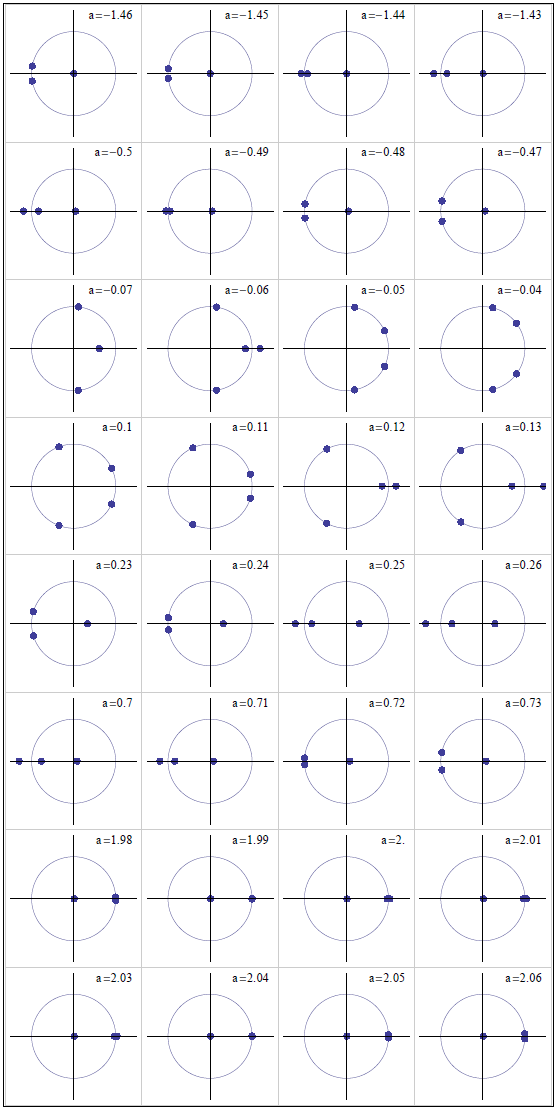}
\caption[]{Similar to Figure~\ref{fig:transitions-0-1}, but for the
  \textit{coupled} system whose stability plot is given in
  Figure~\ref{fig:allmultip-6-4}.  Each row shows the evolution of the
  eigenvalues of the ``mapping at a period'' as one moves along the
  first vertical line (near $q=0.5$) in Figure~\ref{fig:allmultip-6-4},
  near points where the number of stable eigenvalues changes. Such
  changes are represented by changes in the darkness of the $q$-$a$
  area plot in Figure \ref{fig:allmultip-6-4}. The transitions on this
  first vertical line (near $q=0.5$) all have analogues in
  the decoupled case, and the
  ``collisions'' of eigenvalues still happen on the real line only.}\label{fig:transitions-6-4-1}
\end{figure}

\begin{figure}[H]
\centering
\includegraphics[scale=0.4]{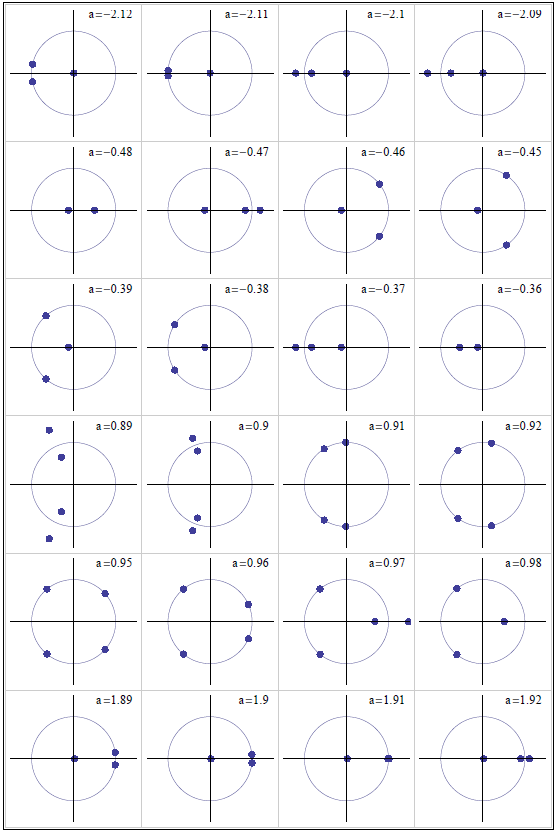}
\caption[]{Similar to Figure \ref{fig:transitions-0-2}, but for the
  \textit{coupled} system of Figure~\ref{fig:allmultip-6-4}.
  Each row shows the evolution of the
  eigenvalues of the ``mapping at a period'' as one moves along the
  second vertical line (near $q=1.5$) in Figure \ref{fig:allmultip-6-4},
  near points where the number of stable eigenvalues changes. This
  time, we encounter a so-called ``\textit{combined resonance}'' near $a =
  0.9$, which doesn't have an analogue in the decoupled case. This is
  represented as a collision of eigenvalues on the unit circle,
  \textit{away} from the real line.}\label{fig:transitions-6-4-2}
\end{figure}

\section{The infinite determinant method}\label{inf-sec}

The ``infinite determinant'' approach to the stability analysis of
the Mathieu
equation \cite{whittaker1996course}
consists of substituting a modified Fourier expansion
(a Floquet expansion)
into the equations of motion
and obtaining
an infinite set of linear equations for the Fourier coefficients.
These equations have
nontrivial solutions only when the determinant of a certain infinite
rank matrix
(involving the parameters $q$, $a$, $\alpha$, $\theta$) vanishes. In practice, the
infinite determinant is replaced by the determinant of a large-rank,
finite matrix, and setting this determinant to zero gives
approximate results.
We will describe this method in the setting of the
single variable Mathieu equation, explain how it can be generalized to
the multi-variable case, and present plots
for certain ``simple'' stability boundaries of the coupled system
that can be obtained from this method in a straightforward way. These boundaries
are in
agreement with the results of the numerical approach described in the
previous section.

Using Floquet's theorem, we look for a
solution to,
\begin{equation}\label{single-mathieu-for-series}
  \ddot{x} + (a + 2q\cos{2\tau})x=0\,,
\end{equation}
of the form,
\begin{equation}\label{single-var-series}
  x(\tau) = e^{i\nu \tau}\sum_{n=-\infty}^{\infty} b_n e^{i2n\tau}\,.
\end{equation}
Here, the infinite sum represents a periodic function with
period equal to
the period of the sinusoidal term in (\ref{single-mathieu-for-series}),
and the
exponential term in front determines the long time stability of the
solution. The system is stable when $\nu$ is purely real, and unstable
when $\nu$ has a nonvanishing imaginary part.\footnote{Positive
  imaginary parts result in decaying solutions, however, our
  discussion in Section~\ref{section:numerical} implies that such solutions are accompanied
  with solutions which have $\nu$ values with negative imaginary parts,
  i.e., those that grow unboundedly in time.}
Substituting (\ref{single-var-series}) into
(\ref{single-mathieu-for-series}) and shifting the summation index in
two of the terms, we get,
\begin{equation}\label{single-var-series-eq}
  e^{i\nu\tau} \sum_{n=-\infty}^{\infty}\left[(-(\nu + 2n)^2 + a)b_n +
    q(b_{n-1} + b_{n+1})\right]e^{i2n\tau} = 0\,.
\end{equation}
Setting the coefficient of each basis functions $e^{i2n\tau}$ to zero,
we obtain an infinite set of equations for the
coefficients $b_n$. This set of equations has a solution only if the
determinant of the infinite matrix of coefficients vanishes, i.e.,
$\det{\mathbf{B}}=0$, where,
\begin{equation}\label{b-matrix}
  B_{mn} = \Bigg\{
    \begin{array}{ll}
      -(\nu+2n)^2 + a & : m=n\\
      q               & : m=n\pm 1\\
      0               & : \mbox{Otherwise.}
    \end{array}
\end{equation}
The determinant of the matrix (\ref{b-matrix}) does not
converge as it stands,\footnote{We could remedy this by
obtaining an alternative set of
equations equivalent to (\ref{single-var-series-eq}) by
dividing each equation derived from (\ref{single-var-series-eq})
by the corresponding diagonal entry, $B_{nn}$.
Working with the matrix resulting from such an alternative set
of equations results
in a convergent determinant.}
however, in order to extract numerical results,
we will work with a finite-size
version of the matrix, and the
convergence issue will not be of practical concern for the sizes and the numerical precision
we will be working with.

Setting the determinant of $\mathbf{B}$ to zero
gives an equation that relates $q$, $a$, and the
exponent $\nu$. Once again, for given $q$ and $a$, we can extract
the growth factor, this time by solving the resulting algebraic
equation.
Checking whether the equation for $\nu$ has an imaginary
solution allows us to
deduce the stability/instability of the system for the given values of
$q$ and $a$.
Looping over values of $q$ and $a$ and repeating the analysis
would give the relevant
stability plots. For the single variable case (or for the decoupled,
two-variable case), it is possible to extract
 the stability boundaries \textit{directly}, without looping
 over
 $q$ and $a$ and finding the transitions from stability to
 instability---it turns out \cite{hansen1985stability} that for this case,
 setting the
growth factor $e^{i\nu\pi}$ to $\pm 1$ gives
equations that relate $q$
and $a$ on the stability boundaries.

The generalization of the infinite determinant approach to the
multi-variable case is
straightforward, and involves replacing various scalars by
vectors/matrices. 
One still gets a determinant equation relating $a$, $q$ and
$\nu$, but in this case, not all stability boundaries can be
obtained directly, without looping over $q$ and $a$.
Setting $e^{i\nu\pi}$ to $\pm 1$ as in the single variable case
gives the stability boundaries that are related to
the so-called ``natural
resonances'', but there is another set of stability boundaries, namely,
those related to
``combined resonances'', which cannot be obtained by this method. In
\cite{hansen1985stability}, a pragmatic but unrigorous approach is
proposed
for obtaining the
boundaries corresponding to the combined resonances,
but we do not follow this procedure
here. Instead, we only obtain the boundaries that are related to
natural resonances, and compare the results with those we get from the
numerical approach described in the previous section.

\begin{figure}[H]
\centering
\includegraphics[scale=0.7]{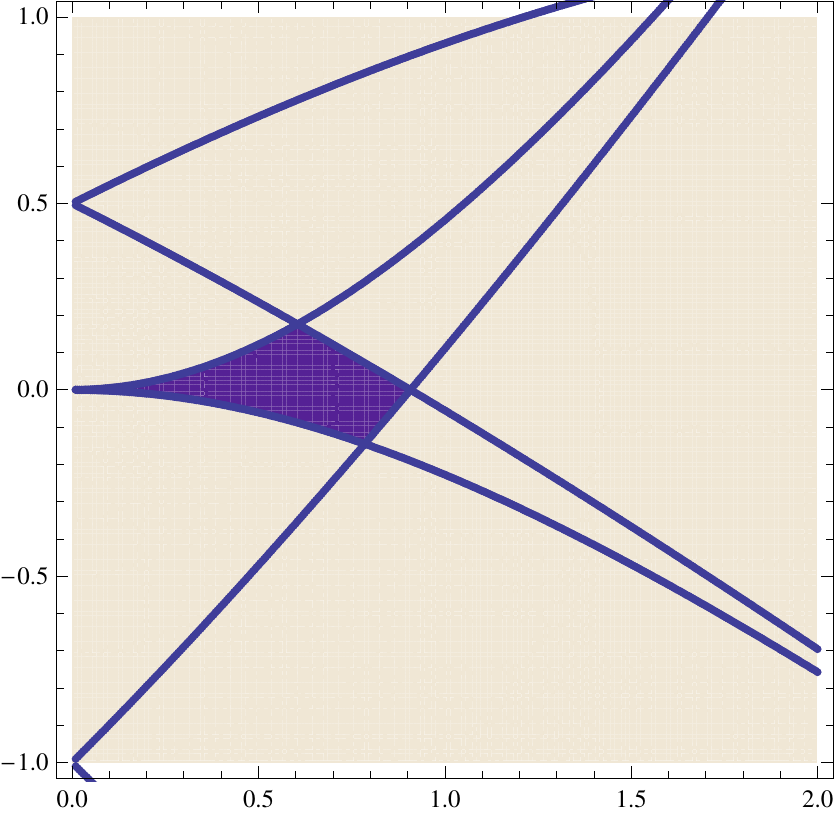}
\caption[]{Stability boundaries obtained from the infinite determinant
method (the curves) shown together with the primary stability region
obtained by
numerical analysis (dark region), for the
decoupled, $\theta=0^{\circ}$ case, and $\alpha=1/2$.
The curves obtained from the infinite
determinant method indeed form boundaries for the primary stability region.}
\label{fig:infinite_det_theta_0}
\end{figure}

\begin{figure}[H]
\centering
\includegraphics[scale=0.7]{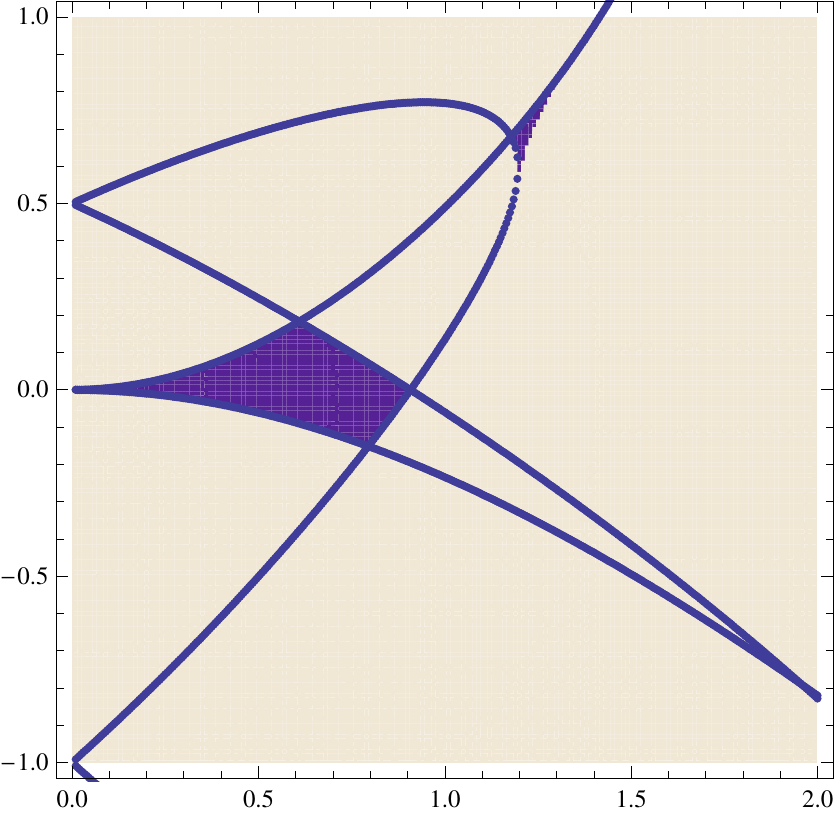}
\caption[]{Stability boundaries obtained from the infinite determinant
method (curves) shown together with the stability regions obtained by
numerical analysis (dark regions), for
$\theta=12^{\circ}$, $\alpha=1/2$. The curves give all the boundaries of
the primary stability
region, but not all the boundaries of a small, secondary
stability region.}\label{fig:infinite_det_theta_12}
\end{figure}

\begin{figure}[H]
\centering
\includegraphics[scale=0.7]{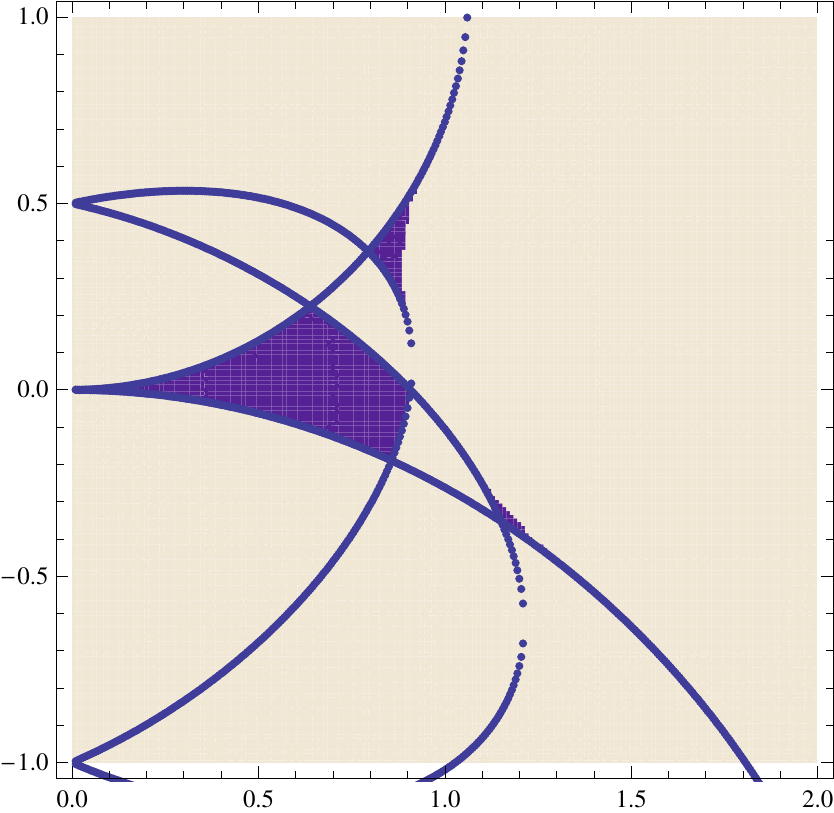}
\caption[]{Similar to Figures \ref{fig:infinite_det_theta_0} and \ref{fig:infinite_det_theta_12},
for
$\theta=32^{\circ}$. As in Figure \ref{fig:infinite_det_theta_12},
the curves obtained from the infinite
determinant method form boundaries for the primary stability
region, but do not give all the boundaries of two small, secondary
stability regions.}\label{fig:infinite_det_theta_32}
\end{figure}

\begin{figure}[H]
\centering
\includegraphics[scale=0.7]{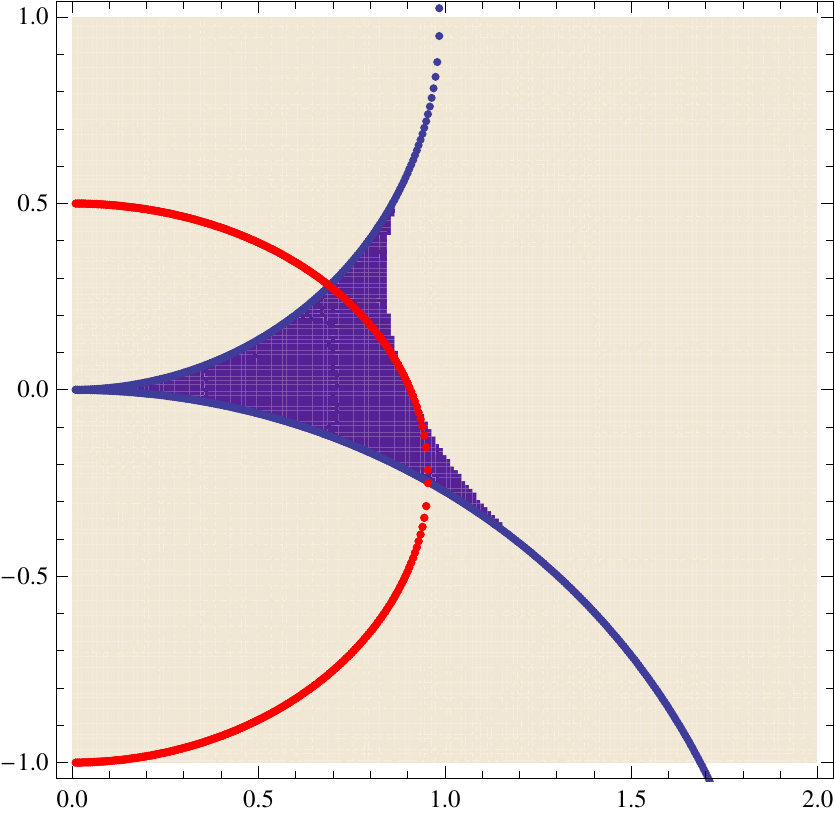}
\caption[]{Similar to Figures \ref{fig:infinite_det_theta_0} and \ref{fig:infinite_det_theta_12},
\ref{fig:infinite_det_theta_32},
for
the special angle $\theta=45^{\circ}$. The red curves obtained from the infinite
determinant method  fail to give the boundaries of the enhanced
stability region, which is formed by the joining of the primary stability region
with the secondary regions. See the following sections
for more on the behavior around this special angle.}\label{fig:infinite_det_theta_45}
\end{figure}

In Figures
\ref{fig:infinite_det_theta_0}-\ref{fig:infinite_det_theta_45},
we give the regions of complete stability obtained by the
methods of the previous
section, and the stability boundaries due to the ``natural
resonances'',
obtained by the infinite determinant method described here.
We choose $\alpha=1/2$, and show plots for 4 different values of
$\theta$. As can be seen, the boundaries obtained by
the infinite determinant method bound the primary stability region
accurately, except for the special case of
$\theta=45^{\circ}$.\footnote{We will give a more comprehensive set of
stability plots in the following sections, where the behavior around
$\theta=45^{\circ}$ will become more transparent.}
Smaller secondary stability regions, which move as the
angle $\theta$ changes, also have two boundaries that are given
accurately by the infinite determinant analysis. However,
these regions also have a third boundary that is invisible to the method
employed here, which is capable of getting only the boundaries due to
natural resonances.

\section{Method of multiple scales}\label{section:multiple}
For a given set of parameters, the Floquet analysis of Section~\ref{numstab}
gives the stability regions of the coupled equations of motion
numerically. While such results are useful, for asymmetric trap design
and characterization, it would be beneficial to have at least
approximate analytical formulae for the stability boundaries. Such
approximate formulae would enable trap designers to quickly check the stability
regions for a particular trap geometry, and would guide the trap
design. Analogous results for the case of symmetric traps are
well-known and useful.

One possible approach to this problem is to use straightforward
perturbation theory of ordinary differential equations, which involves
treating a certain parameter ($q$, in our case) as small, and
expanding the unknown solution into a power series in terms of this
parameter. This gives separate equations for each power of the
perturbation parameter, which are then solved order by order.  While
useful in many settings, this approach fails for a variety of problems
due to the appearance of so-called ``secular'', or ``resonant'' terms,
which grow in time in an unbounded manner even when the exact solution
is known to be bounded. This behavior invalidates the assumptions in
the perturbation analysis after a certain time period. A
straightforward perturbation expansion of the Mathieu equation
results in such behavior.

Multi-scale perturbation theory, or the method of multiple scales, is
a more sophisticated technique capable of avoiding such secular terms
by expanding not only the solution, but also the independent parameter
(typically, time) to a series in the small parameter.  This results in
a set of \textit{partial} differential equations (PDEs) that
collectively represent the ordinary diffential equation under
consideration.  The solutions of these PDEs are then restricted by
imposing the condition that secular terms do not appear. The absence
of secular terms allows the method of multiple scales to give
uniformly accurate results for cases where standard perturbation
theory gives results valid only for short time intervals. In this
Section, we will use this technique to obtain approximate
stability boundaries for the coupled system,
(\ref{eomfin1})-(\ref{eomfin2}). Since the calculations are a bit tedious,
for readers only interested in the final results, we note
that the approximate formulas for the stability boundaries are
given in Equations (\ref{doublea0}), (\ref{doublea0neg}), (\ref{doublea1})
and (\ref{doubleanegative}).

The method of multiple scales has been applied to a wide variety of
problems in physics, engineering, and applied mathematics. There are
various conventions followed in the literature; below, we will use the
approach described in \cite{nayfeh1973perturbation} and
\cite{nayfeh1979nonlinear}.  For an elementary introduction to the
method of multiple scales, see
\cite{kevorkian,nayfeh1973perturbation}. For applications to the
coupled Mathieu system, see \cite{nayfeh1979nonlinear,
mahmoud1997stability}.  Note that the analyses in
\cite{nayfeh1979nonlinear} and \cite{mahmoud1997stability}
are not immediately
applicable to our case, since, as opposed to the cases covered in
these references, some of the natural frequencies for the Paul traps
we are considering are imaginary (i.e., one of the transverse
directions is anti-confining).

\subsection{Single-variable Mathieu equation}\label{sec:single-mathieu}
Let us first demonstrate the use of multi-scale perturbation theory
on the relatively simple case of the single-variable Mathieu equation.
After this example,
we will work out the multi-scale perturbation analysis of the
coupled system (\ref{eomfin1})-(\ref{eomfin2})
describing asymmetric surface traps.

\paragraph{Preliminary discussion.}
We would like to obtain approximate formulas for $a(q)$ on the stability
boundaries of the Mathieu equation,
\begin{equation}\label{single-mathieu}
    \ddot{x} + (a + 2q\cos 2\tau)x = 0\,.
\end{equation}
This equation has multiple regions of stability separated by regions of
instability; we will focus on the primary stability region
near the origin of the $a$-$q$ plane. We will treat $q$ as a small
variable, and perform an expansion in its powers.

We begin the multi-scale perturbation analysis by introducing a set of
slowly-varying time variables (the ``multiple scales''),
\begin{equation}
  T_0 = \tau,\,\,T_1 = q\tau,\,\,T_2 = q^2\tau,\ldots\,.
\end{equation}
We replace the dependent variable $x$ with a function of the $T_i$s,
\begin{equation}\label{x-multi}
  x(\tau) = x(T_0, T_1, \ldots)\,,
\end{equation}
and expand it into a series in $q$,
\begin{equation}\label{x-expand}
  x(T_0,T_1,\ldots) = x_0(T_0,T_1,\ldots) + qx_1(T_0,T_1,\ldots) +
    q^2x_2(T_0,T_1,\ldots) + \ldots\,.
\end{equation}
The chain rule gives the derivative with respect to $\tau$ in terms of partial derivatives
with respect to the time variables $T_i$,
\begin{equation}\label{dt-expand}
  \frac{d}{d\tau} = \frac{\partial}{\partial T_0} +
  q \frac{\partial}{\partial T_1} +
  q^2 \frac{\partial}{\partial T_2} + \ldots
\end{equation}
Substituting  (\ref{x-multi}),
(\ref{x-expand}) and (\ref{dt-expand}) into (\ref{single-mathieu}),
one gets the equations for multi-scale perturbation theory.
These equations give approximate solutions
to (\ref{single-mathieu}) that are valid for small values of $q$, unless $a$ is
close to certain critical values. These critical values are of special
interest to us, since they are the locations where
the stability boundaries
intersect the $q=0$ axis. Various approaches exist for dealing with
these critical points; we will follow the technique used in
\cite{nayfeh1973perturbation},
where one expands $a$ into a series, as well, around a critical
point $a_0$,
\begin{equation}\label{a-expand}
 a = a_0 + a_1 q + a_2 q^2 + \ldots\,.
\end{equation}
In this approach, one obtains the stability boundaries by first
getting
conditions on $a_1$, $a_2,\ldots$ that ensure that the approximate
solutions are bounded. These conditions give the
critical values of these coefficients, and substituting these critical values
back into (\ref{a-expand}), one gets the stability
boundaries. It is noted in~\cite{nayfeh1979nonlinear} that in order to
get results accurate to order $q^2$, it suffices to work with $T_0$
and $T_1$. Accordingly, we will ignore $T_2$ and higher order ``scales''
and let $x(\tau) = x(T_0, T_1)$.

\paragraph{The equations.}
Using the notation $D_i = \frac{\partial}{\partial T_i}$, we have, up
to second order in $q$,
\begin{eqnarray}\label{d2}
  \frac{dx}{d\tau} &=& D_0x_0 + q (D_1x_0 + D_0x_1) + q^2(D_1x_1 + D_0x_2) \\
\label{d22}  \frac{d^2x}{d\tau^2} &=& D_0^2x_0 + q(D_0^2x_1 + 2D_0D_1x_0) +
  q^2(D_1^2x_0 + 2D_0D_1x_1 + D_0^2x_2)\,.
\end{eqnarray}
Substituting (\ref{d2}), (\ref{d22}) and (\ref{a-expand}) (where $a_0$
is arbirary, for now) in (\ref{single-mathieu}), we get,
\begin{align*}
    D_0^2x_0 &+ q (D_0^2x_1 + 2D_0D_1x_0) + q^2 (D_1^2x_0+2D_0D_1x_1+D_0^2x_2)\\
      &+(a_0 + q a_1 + q^2 a_2)(x_0+q x_1+q^2 x_2) +
    2q (x_0+q x_1+q^2 x_2)\cos2T_0=0\,.
\end{align*}
Collecting terms with similar powers of $q$ and setting them to zero,
we get the following partial differential equations:
\begin{eqnarray}\label{x0s}
  D_0^2 x_0 + a_0 x_0 &=& 0 \\ \label{x1s}
  D_0^2 x_1 + a_0 x_1 &=& -2D_0D_1x_0 - a_1x_0 - 2x_0\cos{2T_0} \\
  D_0^2 x_2 + a_0 x_2 &=& -2D_0D_1x_1 - D_1^2x_0 -a_1 x_1 -
    a_2 x_0 -2x_1 \cos{2T_0}\label{x2s}
\end{eqnarray}
As mentioned above, if the value of $a$ under consideration is away
from a set of critical
values, one can ignore $a_1$ and $a_2$, set
$a=a_0$, and solve the equations (\ref{x0s})-(\ref{x2s}) to obtain an
approximate solution to (\ref{single-mathieu}). This approach, in fact,
is how the critical points are obtained in the first place: we set $a=a_0$, and
obtain the special values of $a_0$ for which the approach fails (due to
the appearence of ``small denominators''). See, e.g.,
\cite{nayfeh1973perturbation,nayfeh1979nonlinear}
for a discussion and examples.

Here, we will skip this step, and just borrow the result that
$a_0=0$
and $a_0=1$ are the two critical values relevant for the first
stability region near the origin. We next perform expansions around
these two points.

\subsubsection{Expansion around $a=0$}\label{sec:single-mathieu-a0}
Setting $a_0 = 0$ in (\ref{x0s})-(\ref{x2s}), we get,
\begin{align}
   D_0^2x_0 &= 0\label{x0n} \\
   D_0^2x_1 &= -2D_0D_1x_0 - a_1 x_0 - 2x_0 \cos{2T_0} \label{x1n}\\
   D_0^2x_2 &= -2D_0D_1x_1 - D_1^2x_0 - a_2 x_0 - a_1x_1 -
     2x_1 \cos{2T_0}\,. \label{x2n}
\end{align}
The general solution to (\ref{x0n}) is,
\begin{equation}\label{zerothxsolgen}
  x_0(T_0,T_1)=A(T_1)+T_0 B(T_1)\,,
\end{equation}  where $A$ and $B$ are arbitrary functions.  In order to
suppress the secular term, i.e., the term that grows in time, we set
$B=0$. This gives,
\begin{equation}\label{zerothxsol}
  x_0(T_0,T_1) = A(T_1)\,.
\end{equation}
Substituting (\ref{zerothxsol}) into (\ref{x1n}), we get,
\begin{equation}\label{x1eqsubs}
  D_0^2x_1 = A_0(T_1) (-a_1 - 2\cos{2T_0})\,.
\end{equation}
The solution for $x_1$ has a secular term
proportional to $a_1$; in order to suppress unbounded growth, we set
$a_1=0$.  Solving (\ref{x1eqsubs}) with this assumption and setting to
zero another secular term that appears (this term is analogous to the
$B$ term in (\ref{zerothxsolgen})), we get,
\begin{equation}\label{x1sol}
  x_1(T_0,T_1)=C(T_1)+\frac{1}{2} A(T_1) \cos{2T_0}\,.
\end{equation}
Substituting (\ref{zerothxsol}) and (\ref{x1sol}) in (\ref{x2n}) and
using $a_1=0$, we get,
\begin{align*}
    D_0^2x_2=&2A'(T_1)\sin{2T_0}-A''(T_1)-a_2 A(T_1)-2 C(T_1) \cos{2T_0}\\
    &-\frac{A(T_1)}{2}-\frac{A(T_1)}{2} \cos{4T_0}\,.
\end{align*}
In order to avoid terms in $x_2$ that grow linearly in $T_0$, we must have,
\begin{equation*}
  A''(T_1)+(a_2+ \frac{1}{2})A(T_1) = 0\,.
\end{equation*}
This equation will have non-growing solutions for $A(T_1)$ if and only if,
\begin{equation*}
   a_2 \ge -\frac{1}{2}\,.
\end{equation*}
Combining the conditions $a_0=0$, $a_1=0$, and $a_2\ge -1/2$,
we see that the curve in the $a$-$q$ plane separating
bounded solutions from unbounded ones around
$a=0$, $q=0$ is given by,
\begin{equation}\label{singlea0}
   a = -\frac{1}{2}q^2\,,
\end{equation}
with $a > -q^2/2$ being the stable region.


\subsubsection{Expansion around $a = 1$}

Setting $a_0=1$ in (\ref{x0s})-(\ref{x2s}), we get,
\begin{eqnarray}\label{x0s1}
  D_0^2 x_0 + x_0 &=& 0 \\ \label{x1s1}
  D_0^2 x_1 + x_1 &=& -2D_0D_1x_0 - a_1x_0 - 2x_0\cos{2T_0} \\\label{x2s1}
  D_0^2 x_2 + x_2 &=&
    -2D_0D_1x_1 - D_1^2x_0 -a_1 x_1 - a_2 x_0 -2x_1\cos{2T_0}\,.
\end{eqnarray}
The general solution of (\ref{x0s1}) is,
\begin{equation}
  x_0(T_0,T_1)=A(T_1) \cos{T_0} + B(T_1) \sin{T_0}\,.
\end{equation}
Plugging this in (\ref{x1s1}), we get,
\begin{equation}\label{g}
  \begin{split}
    D_0^2x_1+x_1=&2A'\sin{T_0} - 2B'\cos{T_0}-a_1(A\cos{T_0}+B\sin{T_0})\\
      &-A(\cos{T_0}+\cos{3T_0})+B(\sin{T_0}-\sin{3T_0})\,.
  \end{split}
\end{equation}
The solutions of this equation will have growing parts unless the
coefficients of the ``resonance terms'' $\sin{T_0}$ and $\cos{T_0}$
on the right hand side vanish. This gives,
\begin{eqnarray}
  2A'+(1-a_1)B&=&0\label{a0}\\
  2B'+(1+a_1)A&=&0\label{b0}\,.
\end{eqnarray}
This system will have exponentially growing solutions if $a_1^2<1$,
and oscillatory solutions if $a_1^2>1$. Thus, the critical values of
$a_1$ are $a_1 = \pm 1$. Assuming $a_1^2>1$, the general solution is,
\begin{eqnarray}\label{sola0}
  A(T_1) &=& c \sin{\lambda T_1} +  d \cos{\lambda T_1}\\
  B(T_1) &=& \frac{2\lambda}{a_1-1}(-d\sin{\lambda T_1} + c\cos{\lambda T_1})\,,\label{solb0}
\end{eqnarray}
where $\lambda=\sqrt{\frac{a_1^2-1}{4}}$ and $c$ and $d$ are
constants. After substituting (\ref{sola0})-(\ref{solb0}) in (\ref{g}) and
setting the resulting resonant terms to zero, we get,
\begin{equation}
  D_0^2x_1+x_1 = -(A \cos{3T_0} + B \sin{3T_0})\,,
\end{equation}
whose general solution is,
\begin{equation}\label{x1fora1}
  \begin{split}
    x_1(T_0,T_1) =& C(T_1)\cos{T_0} +D(T_1)\sin{T_0} \\
      &+\frac{A(T_1)}{8}  \cos{3T_0}+ \frac{B(T_1)}{8}  \sin{3T_0}\,.
  \end{split}
\end{equation}
Plugging (\ref{x1fora1}) in (\ref{x2s1}), the right hand side of
(\ref{x2s1}) ends up having various terms proportional to $\sin{T_0}$
and $\cos{T_0}$. Once again, these resonant terms will result in
growing solutions for $x_2$, so we set their coefficients to zero.
This gives,
\begin{eqnarray}
  2C'+(1-a_1)D &=& B'' + B(a_2+\frac{1}{8})\\
  2D'+(1+a_1)C &=& -A''-A(a_2+\frac{1}{8})\,.
\end{eqnarray}
Recalling $A''=-\lambda^2A$ and $B''=-\lambda^2B$ from above (see
(\ref{sola0}-\ref{solb0})), we get,
\begin{eqnarray}
  2C'+(1-a_1)D &=&  B \big(\frac{1-a_1^2}{4} + a_2+\frac{1}{8}\big)\\
  2D'+(1+a_1)C &=& -A \big(\frac{1-a_1^2}{4} + a_2+\frac{1}{8}\big)\,.
\end{eqnarray}
Combining these equations, using the explicit solution
(\ref{sola0}-\ref{solb0})
for $A$ and $B$ and once again requiring the absence of resonant
terms, we get,
\begin{equation}
  a_2 = -\frac{1}{8} + \frac{a_1^2-1}{4}\,.
\end{equation}
Finally, combining the conditions for $a_0$, $a_1$ and $a_2$, we get
the formula for the stability boundaries near $q=0$, $a=1$ as,
\begin{eqnarray}
  a &=& a_0 + a_1 q + a_2q^2\\
   &=& 1\pm q-\frac{1}{8}q^2\,,\label{singlea1}
\end{eqnarray}
with the region between the two curves corresponding to unstable
solutions, and the region outside corresponding to stable ones. The results
(\ref{singlea0}) and (\ref{singlea1}) are in agreement with the classical
results for Mathieu equation.

For future reference, we note that if the $a$ in the Mathieu equation
(\ref{single-mathieu}) was replaced
with $-\alpha a$, the stability boundaries
(\ref{singlea0}) and (\ref{singlea1}) would be replaced with,
\begin{equation}
   a = \frac{-1}{\alpha}\left(1\pm q-\frac{1}{8}q^2\right)\,,\label{singlea1alpha}
\end{equation}
and ,
\begin{equation}\label{singlea0alpha}
   a = \frac{1}{2\alpha}q^2\,.
\end{equation}

\subsection{Two-variable, coupled Mathieu's equations}
We next turn to the coupled Mathieu system
(\ref{eomfin1})-(\ref{eomfin2}), which we reproduce here:
\begin{eqnarray}
  \ddot{x} + ax + 2q(cx+sy)\cos{2\tau} &=& 0\label{xeom2}\\
  \ddot{y} - \alpha ay + 2q(sx-cy)\cos{2\tau} &=& 0\,.\label{yeom2}
\end{eqnarray}
These equations describe the radial motion of an ion in an
asymmetric trap with decoupled axial motion. Recall that the
coordinates are
chosen to be along the DC principal axes, and the constants $c$ and
$s$ are given in terms of the relative angle $\theta$ between the RF
and DC axes as,
\begin{eqnarray}
  c &=& \cos{2\theta}\label{cosc}\\
  s &=& \sin{2\theta}\label{sins}\,.
\end{eqnarray}

As in the single variable case, we begin the multi-scale analysis by
promoting $x$ and $y$ to functions of two separate time variables,
$T_0 = \tau$ and $T_1 = q\tau$, and expanding them to series in $q$,
\begin{eqnarray}
  x &=& x_0(T_0, T_1) + q x_1(T_0, T_1) + q^2 x_2(T_0, T_1) + \ldots\\
  y &=& y_0(T_0, T_1) + q y_1(T_0, T_1) + q^2 y_2(T_0, T_1) + \ldots \,.
\end{eqnarray}
If $a$ is near a critical value $a_0$ where this
expansion fails to work, we expand $a$ as well,
\begin{equation}
  a=a_0+a_1q+a_2q^2+\ldots\,.
\end{equation}
Working to second order in $q$, and collecting terms with similar
powers of $q$, we get,
\begin{align}\label{x0k}
  D_0^2 x_0 + a_0 x_0 =& 0 \\ \label{x1k}
  D_0^2 x_1 + a_0 x_1 =& -2D_0D_1x_0 - a_1x_0 - 2(cx_0+sy_0)\cos{2T_0} \\ \label{x2k}
  \begin{split}
  D_0^2 x_2 + a_0 x_2 =& - 2D_0D_1x_1 - D_1^2x_0 -a_1 x_1 \\
  &\quad{}- a_2 x_0 -2(c x_1+s y_1)\cos{2T_0}
  \end{split}
\end{align}
for (\ref{xeom2}), and,
\begin{align}\label{y0k}
  D_0^2 y_0 -\alpha a_0 y_0 =& 0 \\ \label{y1k}
  D_0^2 y_1 -\alpha a_0 y_1 =& -2D_0D_1 y_0 +\alpha a_1y_0-2(sx_0-cy_0)
    \cos{2T_0}\\ \label{y2k}
  \begin{split}
   D_0^2 y_2 -\alpha a_0 y_2 =& - 2D_0D_1 y_1 - D_1^2y_0 + \alpha a_1 y_1 \\
   &\quad{}+ \alpha a_2 y_0 -2(sx_1-cy_1)\cos{2T_0}
  \end{split}
\end{align}
for (\ref{yeom2}). Following our discussion
in Section~\ref{numstab}, we will assume $\alpha>0$. For $a\ge 0$,
the critical values are $a=0$ and $a=1$ (once again we will skip the
justification for these values, and refer the reader to
\cite{nayfeh1979nonlinear}).
The expansion around $a=0$ will proceed similarly to the single
variable case, but the $a=1$ case requires special attention. Assuming
$a_0>0$ and $\alpha>0$, we see that the solutions of (\ref{x0k}) are
oscillatory, but the ones for (\ref{y0k}) are exponential.

Assuming that the initial conditions of the
system are fine-tuned so that the unbounded solutions are not excited,
one can investigate the bounded part of the general solution. However,
as mentioned in our discussion in Section~\ref{numstab}, noise will
undoubtedly excite the unstable solutions in practice.

Nevertheless, our stability plots obtained by the techniques
of Section~\ref{numstab} suggest that the curves separating
regions of partial stability from those of full instability may in
fact be of some use. We observe from the stability plots
of Figures~\ref{fig:allmultip-0} and
\ref{fig:allmultip-6-4} (and from additional plots
we will present below) that
the primary stability region around the origin is
bounded by an extension of such curves. In other words, the curves
emanating at the critical point $a_0=1$ begin as
boundaries between regions of full instability
and regions of partial
instability, but once they intersect with the
stability boundaries emanating from $a_0=0$, they become
boundaries between full instability and full stability.
Motivated by this
empirical observation, we will obtain formulas for the curves emanating at
the critical point $a_0=1$, and use them as stability boundaries after
they intersect with the curves emanating from $a_0=0$. We will
demonstrate the reliability of this approach in the stability plots we present
below.

\subsubsection{Expansion around $a = 0$}
We begin with the expansion around $a=0$.
Setting $a_0=0$ in equations (\ref{x0k})-(\ref{y2k}), we get,
\begin{align}
D_0^2x_0 =& 0\label{da0x0}\\
D_0^2x_1 =& -2D_0D_1x_0-a_1x_0-2 c x_0 \cos{2T_0} - 2 s y_0 \cos{2T_0}\label{da0x1}\\
\begin{split}
D_0^2x_2 =& -2D_0D_1x_1-D_1^2x_0-a_1x_1-a_2x_0-2c x_1\cos{2T_0}\\
   &\qquad{}- 2s y_1\cos{2T_0} \label{da0x2}\,,
\end{split}
\end{align}
and,
\begin{align}
D_0^2y_0 =& 0\label{da0y0}\\
D_0^2y_1 =& -2D_0D_1y_0+\alpha a_1y_0 + 2c y_0 \cos{2T_0}  - 2s x_0 \cos{2T_0}
  \label{da0y1}\\
\begin{split}
D_0^2y_2 =& -2D_0D_1y_1 - D_1^2y_0 +\alpha a_1y_1+ \alpha a_2 y_0 +
  2c y_1\cos{2T_0} \\
  &\qquad{}-2s x_1\cos{2T_0}\,. \label{da0y2}
\end{split}
\end{align}
Solving (\ref{da0x0}) and (\ref{da0y0}) and setting the secular terms
to zero, we get,
\begin{eqnarray}
  x_0(T_0,T_1)&=&A(T_1)\label{x0sol}\\
  y_0(T_0,T_1)&=&C(T_1)\label{y0sol}\,.
\end{eqnarray}
Substituting these in (\ref{da0x1}) and (\ref{da0y1}) gives,
\begin{eqnarray}
  D_0^2x_1 &=&       - a_1A(T_1)-2(c A(T_1)+s C(T_1)) \cos{2T_0}\label{x11}\\
  D_0^2y_1 &=&  \alpha a_1C(T_1)+2(c C(T_1)-s A(T_1)) \cos{2T_0}\label{y11}\,.
\end{eqnarray}
In order to avoid growing $x_1$ and $y_1$,
we need to pick $a_1=0$.
Solving (\ref{x11}) and (\ref{y11}) with this assumption and setting
the coefficients of two other secular terms that grow linearly in
$T_0$ to zero, we get,
\begin{eqnarray}
    x_1(T_0,T_1)&=& B(T_1) +
      \frac{1}{2}(cA(T_1)+sC(T_1)\cos{2T_0}\label{x1sol-twovar}\\
    y_1(T_0,T_1)&=& D(T_1) +
      \frac{1}{2}(sA(T_1)-cC(T_1))\cos{2T_0}\label{y1sol-twovar}\,.
\end{eqnarray}
Plugging the solutions (\ref{x0sol}), (\ref{y0sol}),
(\ref{x1sol-twovar}), (\ref{y1sol-twovar}) in (\ref{da0x2}) and
(\ref{da0y2}), we get equations for $x_2$ and $y_2$. These will have
bounded solutions only if the $T_0$-independent terms on the right
hand sides of the equations add up to zero. Asserting this condition gives,
\begin{eqnarray}
  A''(T_1)+\left(a_2         + \frac{1}{2}(c^2+s^2)\right)A(T_1) &=& 0\label{a2forx-1}\\
  C''(T_1)+\left(-\alpha a_2 + \frac{1}{2}(c^2+s^2)\right)C(T_1) &=& 0\label{a2fory-1}\,.
\end{eqnarray}
Equations (\ref{a2forx-1}) and (\ref{a2fory-1}) will have oscillatory
solutions for $A(T_1)$ and $B(T_1)$ when $a_2>-\frac{1}{2}(c^2+s^2)$,
and when $a_2<\frac{1}{2\alpha}(c^2+s^2)$, respectively. Using
(\ref{cosc}) and (\ref{sins}), we have, $s^2+c^2=1$. Assuming
$\alpha>0$, we see that simulatenous stability occurs for
$-\frac{1}{2} < a_2 < \frac{1}{2\alpha}$. In other words, the stability
boundaries around $a=0$ are given by,
\begin{eqnarray}\label{doublea0}
  a &=& -\frac{1}{2}q^2\\
  a &=& \frac{1}{2\alpha}q^2\,.\label{doublea0neg}
\end{eqnarray}
Note that these conditions are what would be obtained from separate
stability analyses of (\ref{xeom2}) and (\ref{yeom2}), respectively,
if the coupling terms in those equations were ignored and
the single-variable results (\ref{singlea0}) and (\ref{singlea0alpha}) of
Section~\ref{sec:single-mathieu-a0}
were used.

\subsubsection{Expansion around $a = 1$}
We next expand around $a=1$. As mentioned above, in this case we will seek
solutions with partial stability.
For $a_0=1$, (\ref{x0k})-(\ref{x2k}) become,
\begin{eqnarray}
\label{da1x0}
  D_0^2 x_0 + x_0 &=& 0 \\
\label{da1x1}
  D_0^2 x_1 + x_1 &=& -2D_0D_1x_0 - a_1x_0 - 2(cx_0+sy_0)\cos{2T_0} \\
\label{da1x2}
  D_0^2 x_2 + x_2 &=& - 2D_0D_1x_1 - D_1^2x_0 -a_1 x_1
  - a_2 x_0 \\
  &&\qquad{}-2(c x_1+s y_1)\cos{2T_0}\,,
\end{eqnarray}
and (\ref{y0k})-(\ref{y2k}) become,
\begin{align}
\label{da1y0}
  D_0^2 y_0 -\alpha y_0 =& 0 \\
  \label{da1y1}
  D_0^2 y_1 -\alpha y_1 =& -2D_0D_1 y_0 +\alpha a_1y_0
   -2(sx_0-cy_0)\cos{2T_0}\\
   \begin{split}
     \label{da1y2}
     D_0^2 y_2 -\alpha y_2 =& - 2D_0D_1 y_1 - D_1^2y_0 + \alpha a_2 y_0 \\
     & \qquad{}+ \alpha a_1 y_1 -2(sx_1-cy_1)\cos{2T_0}\,.
   \end{split}
\end{align}
The general solution to (\ref{da1x0}) is,
\begin{equation}\label{da1x0sol}
  x_0(T_0,T_1)=A(T_1) \cos{T_0} + B(T_1) \sin{T_0}\,.
\end{equation}
Assuming $\alpha>0$, the solution to (\ref{da1y0}) is exponential,
\begin{equation}
  y_0(T_0,T_1) = E(T_1) \exp{(\sqrt\alpha T_0)}
    + F(T_1) \exp{(-\sqrt\alpha T_0)}\,.
\end{equation}
This shows that the general solution to the coupled system is unstable
for $\alpha>0$, $a_0=1$, as discussed above.  In order to investigate partial
stability, we set the coefficients of the exponential solutions to
zero, $E(T_1)=0=F(T_1)$, which gives $y_0(T_0,T_1)=0$.  Substituting
this and (\ref{da1x0sol}) in (\ref{da1x1}), and setting the
coefficients of $\sin{T_0}$ and $\cos{T_0}$ on the right hand side to
zero in order to avoid resonant (secular) terms that would result in
growing solutions for $x_1$, we get,
\begin{eqnarray}
  \label{da1A0}
    2A'-(a_1-c)B &=& 0\\
    \label{da1B0}
    2B'-(a_1+c)A &=& 0\,.
\end{eqnarray}
Combining these, we get,
\begin{eqnarray}\label{appeq}
 A'' &=& -\frac{a_1^2-c^2}{4}A\\
 B'' &=& -\frac{a_1^2-c^2}{4}B\,,\label{bppeq}
\end{eqnarray}
which will have oscillatory solutions if $a_1^2>c^2$ and
exponential ones if $a_1^2<c^2$. Thus, the critical values for partial
stability are,
\begin{equation}\label{a1boundary}
  a_1=\pm c\,.
\end{equation}
The oscillating solutions for $A$ and $B$ are given as,

\begin{eqnarray}\label{sola0two}
  A(T_1) &=& r \sin{\lambda T_1} +  p \cos{\lambda T_1}\\
  B(T_1) &=& \frac{2\lambda}{a_1-c}(-p\sin{\lambda T_1} + r\cos{\lambda T_1})\,,\label{solb0two}
\end{eqnarray}
where $\lambda=\sqrt{\frac{a_1^2-c^2}{4}}$, and $r$ and $p$ are
constants.

After having ensured that the resonant terms in (\ref{da1x1}) vanish
by enforcing (\ref{da1A0}) and (\ref{da1B0}), we can solve
(\ref{da1x1}) to get the oscillatory solutions for $x_1$. We get,
\begin{equation}\label{da1x1sol}
  \begin{split}
    x_1(T_0,T_1) =& C(T_1)  \cos{T_0} + D(T_1) \sin{T_0} \\
    &+ \frac{c}{8} A(T_1) \cos{3T_0}+ \frac{c}{8}B(T_1)\sin{3T_0}\,.
  \end{split}
\end{equation}
Similarly, substituting the solution (\ref{da1x0sol}) in
(\ref{da1y1}), we get a bounded solution for $y_1$ after setting the
coefficients of the exponential terms to zero:

\begin{equation}\label{da1y1sol}\begin{split}
  y_1(T_0,T_1) = &\frac{s}{(\alpha+1)}( A(T_1)\cos{T_0}-B(T_1)\sin{T_0}) \\
    &+\frac{s}{(\alpha+9)}(A(T_1)\cos{3T_0}
  +B(T_1)\sin{3T_0}) \end{split}\,.
\end{equation}
We next substitute (\ref{da1x1sol}) and (\ref{da1y1sol}) into
(\ref{da1x2}) and collect the resonance terms, i.e., terms
proportional to $\sin{T_0}$ and $\cos{T_0}$, on the right hand
side. Setting the coefficients of these resonances to zero in order to avoid
growing solutions for $x_2$, we get,

\begin{eqnarray}
  2C'(T_1) + D(T_1)(-a_1 + c) &=& B(T_1)''+\beta B(T_1)\\
  2D'(T_1) + C(T_1)(a_1  + c) &=&-A''(T_1)-\beta A(T_1)\,,
\end{eqnarray}
where,
\begin{equation}
  \beta = a_2 + \frac{c^2}{8} +
    \frac{2 s^2 (5+\alpha)}{(9+\alpha) (1+\alpha)}\,.
\end{equation}
Using (\ref{appeq})-(\ref{bppeq}), we get,
\begin{eqnarray}\label{cdeq1}
  2C'(T_1) + D(T_1)\left(-a_1 + c\right) &=& \mu B\\
\label{cdeq2}
  2D'(T_1) + C(T_1)\left(a_1 + c\right) &=& -\mu A\,,
\end{eqnarray}

where $\mu = \beta-\lambda^2 = a_2 + \frac{c^2}{8} +
\frac{2s^2(5+\alpha)}{(9+\alpha) (1+\alpha)} - \frac{a_1^2-c^2}{4}$.
Combining (\ref{cdeq1}) and (\ref{cdeq2})
with (\ref{da1A0}) and (\ref{da1B0}), we get the decoupled equations,

\begin{eqnarray}\label{cppeq}
  C''+\frac{a_1^2-c^2}{4}C &=& -\frac{\mu a_1}{2}A\\
  D''+\frac{a_1^2-c^2}{4}D &=& -\frac{\mu a_1}{2}B\,.\label{dppeq}
\end{eqnarray}
Now, since the solutions (\ref{sola0two}) and (\ref{solb0two}) for $A$ and $B$
are in resonance with the left
hand sides of (\ref{cppeq}) and (\ref{dppeq}),
in order to avoid growing solutions,

the
coefficients on the right hand sides of these equations must vanish.
Using (\ref{a1boundary}), this gives,
\begin{equation}
  a_2 = -\frac{c^2}{8} - \frac{2s^2(5+\alpha)}{(1+\alpha)(9+\alpha)}\,.
\end{equation}
Thus,
the second order approximation to the
relevant stability boundary starting at $a=1$ is given as,
\begin{equation}\label{doublea1}
  a = 1 - cq - \left(
  \frac{c^2}{8} + \frac{2s^2(5+\alpha)}{(1+\alpha)(9+\alpha)}
  \right)q^2\,,
\end{equation}
where we picked the negative sign for the first order term
in order to get the curve that approximates the upper
boundary of the primary stability
region.

\paragraph{Boundary of the primary stability region.} In order to
obtain the approximate boundaries of the primary stability region, we
also need to find an approximate formula for the stability boundary
that emanates from negative $a$ when $q=0$. Our expansion around $a=1$
was based on the assumption that $a>0$, so that $x$ had oscillatory
solutions to zeroth order, and $y$ had exponential ones. In order to
obtain the curves for negative $a$, we just replace the roles of $x$
and $y$ by a series of redefinitions. Namely, we set,
\begin{equation}
  \tilde{x} = y\,,~\tilde{y} = x\,,~\tilde{a} = -\alpha
    a\,,~\tilde{\alpha}=1/\alpha\,,~\tilde{c}=-c\,,~\tilde{s}=s\,.\label{var-tfm}
\end{equation}
These redefinitions transform the equations of motion (\ref{xeom2}) and
(\ref{yeom2}) into exactly the same form, with the variables (except
$q$ and $\tau$) being replaced by their tilded counterparts,
\begin{eqnarray}
  \ddot{\tilde{x}} + \tilde{a}\tilde{x}               + 2q(\tilde{c}\tilde{x}+\tilde{s}\tilde{y})\cos{2\tau} &=& 0  \label{tildeq1} \\
  \ddot{\tilde{y}} - \tilde{\alpha}\tilde{a}\tilde{y} + 2q(\tilde{s}\tilde{x}-\tilde{c}\tilde{y})\cos{2\tau} &=& 0  \,.\label{tildeq2}
\end{eqnarray}
Now, if
$a<0$, the transformation (\ref{var-tfm}) makes $\tilde{a}>0$, thus, we can apply
(\ref{doublea1}) to (\ref{tildeq1})-(\ref{tildeq2}) by replacing all the quantities
in (\ref{doublea1}) by their tilded
counterparts. This
gives,
\begin{equation}
  \tilde{a} = 1 - \tilde{c}q - \left(
  \frac{\tilde{c}^2}{8} + \frac{2\tilde{s}^2(5+\tilde{\alpha})}{(1+
    \tilde{\alpha})(9+\tilde{\alpha})}
  \right)q^2\,,
\end{equation}
or, equivalently,
\begin{equation}\label{doubleanegative}
  a = -\frac{1}{\alpha}\Big(1-  c q - \left(
  \frac{c^2}{8} + \frac{2s^2(5+1/\alpha)}{(1+1/\alpha)(9+1/\alpha)}
  \right)\Big)q^2\,.
\end{equation}
This gives the required
approximate boundary of the primary stability region for $a<0$. Note that
the curve starts at $a=-1/\alpha$.

In Figures~ \ref{fig:alpha_0.5}-\ref{fig:alpha_2.5}, we compare the results of this section
with
the results from the numerical analysis of Section~\ref{numstab}, for
various values of $\alpha$ and $\theta$.\footnote{In Figure~\ref{fig:alpha_0.5}, we
do not show the plot  for
$\theta=90^{\circ}$ since it is the same as the plot for $\theta=0^{\circ}$.
In fact, as can be seen from the figure, the stability behavior of the system
is symmetric around $\theta=45^{\circ}$, in the sense that the plots
for $45^{\circ} + \Delta\theta$ and $45^{\circ}-\Delta\theta$ are
identical. This is due to the fact that the equations of motion are
invariant under the transformation $\theta\to \pi/2-\theta$, $y\to
-y$. Since the system is symmetric under reflections of the $y$ axis,
this transformation should leave the stability plot invariant.
In Figures~\ref{fig:alpha_1}-\ref{fig:alpha_2.5}, we only show the stability plots
for $\theta$ between $0^{\circ}$ and
$45^{\circ}$.}
\section{Conclusions and discussion}\label{section:conclusion}

\begin{figure}[H]
\centering
\includegraphics[scale=0.37]{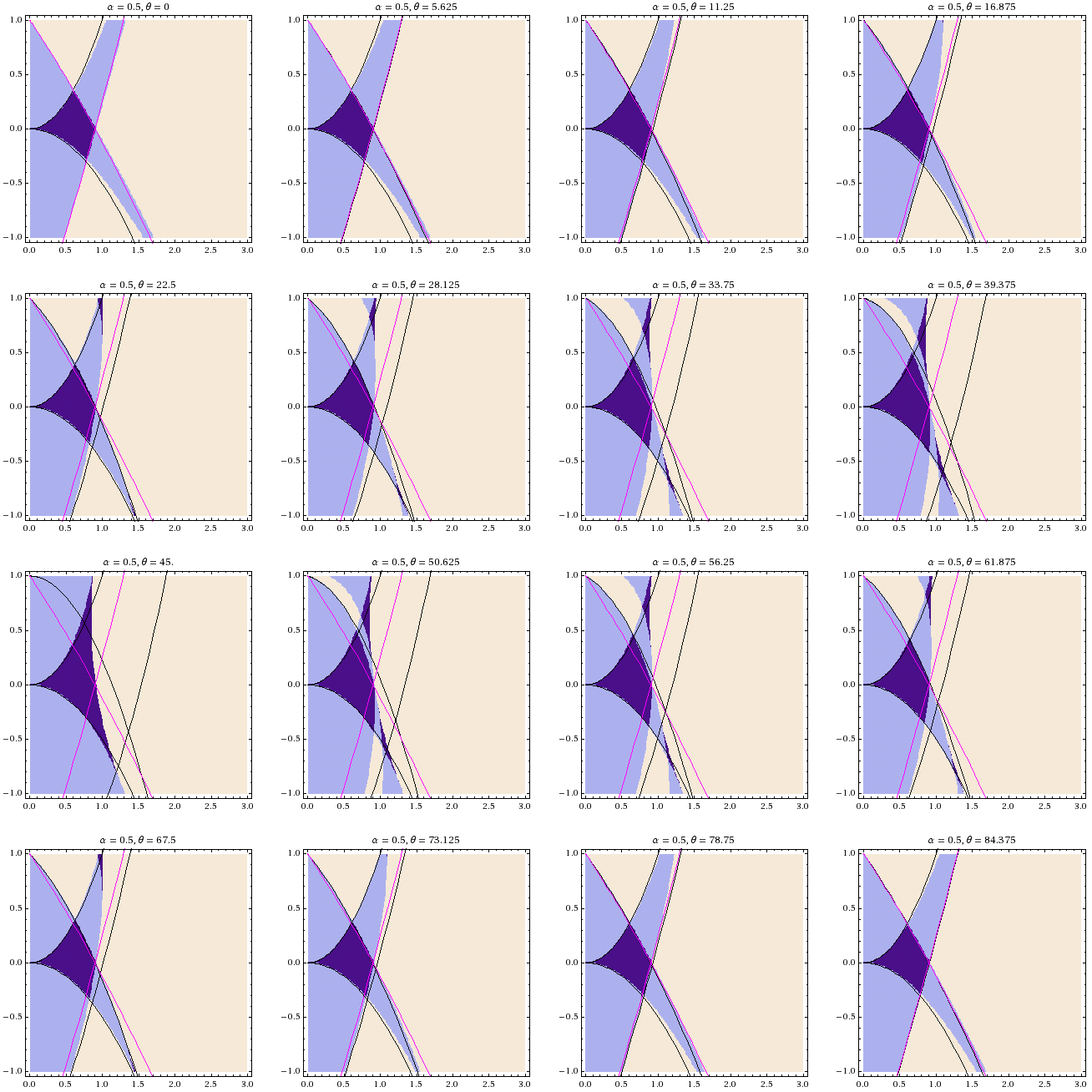}
\caption[]{Stability plots for $\alpha = 0.5$ and $\theta$ between
$0^{\circ}$ and $90^{\circ}$, with a step size of
$5.625^{\circ}$. The $x$-axis is $q$ and the $y$-axis is $a$ in each plot.
The areas shown in dark purple correspond to complete stability,
and those in light purple correspond to partial
stability, predicted using the numerical method of Section~\ref{numstab}.
The black curves are approximate stability boundaries for the coupled,
two variable Mathieu system,
obtained from equations (\ref{doublea0}), (\ref{doublea0neg}), (\ref{doublea1}),
and (\ref{doubleanegative}).
The red curves are the approximate stability
boundaries for the corresponding \textit{decoupled} systems,
obtained from equations
(\ref{singlea1}) and (\ref{singlea1alpha}).
The approximate boundaries around $a=0$
for the coupled and the decoupled systems are identical (see
equations (\ref{singlea0})-(\ref{singlea0alpha}) and (\ref{doublea0})), thus, we only show
black curves for the approximate boundaries passing through $a=0$.
(see the final paragraph of Section
\ref{section:conclusion}
for a discussion of the relation between the
coupled system and the decoupled system).
We see that the black curves follow the partial stability boundaries
quite accurately when $q$ is small, however, when $\theta$ gets
close to $45^{\circ}$, the curves starting at $a=1$ and $a=-1/\alpha = -2$
(the latter point is outside the region shown in the plots)
lose their accuracy near the primary region of full
stability.
The colored curves representing the approximate boundaries of the
\textit{decoupled system}, while inaccurate predictors for
the coupled system,
consistently
underestimate the size of the primary stability region.}\label{fig:alpha_0.5}
\end{figure}

\begin{figure}[H]
\centering
\includegraphics[scale=0.4]{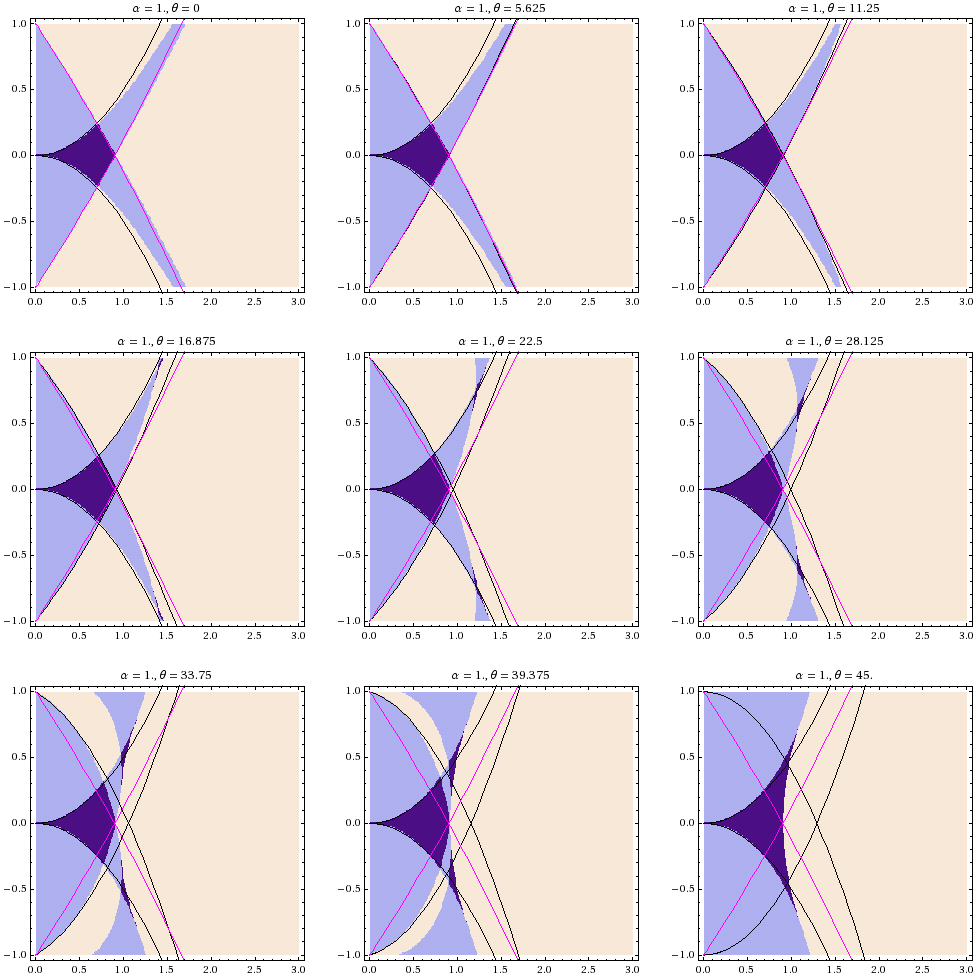}
\caption[]{Stability plots for $\alpha = 1.0$ and $\theta$ between
$0^{\circ}$ and $45^{\circ}$, with a step size of
$5.625^{\circ}$. See the caption for
Figure~\ref{fig:alpha_0.5}.}\label{fig:alpha_1}
\end{figure}

\begin{figure}[H]
\centering
\includegraphics[scale=0.4]{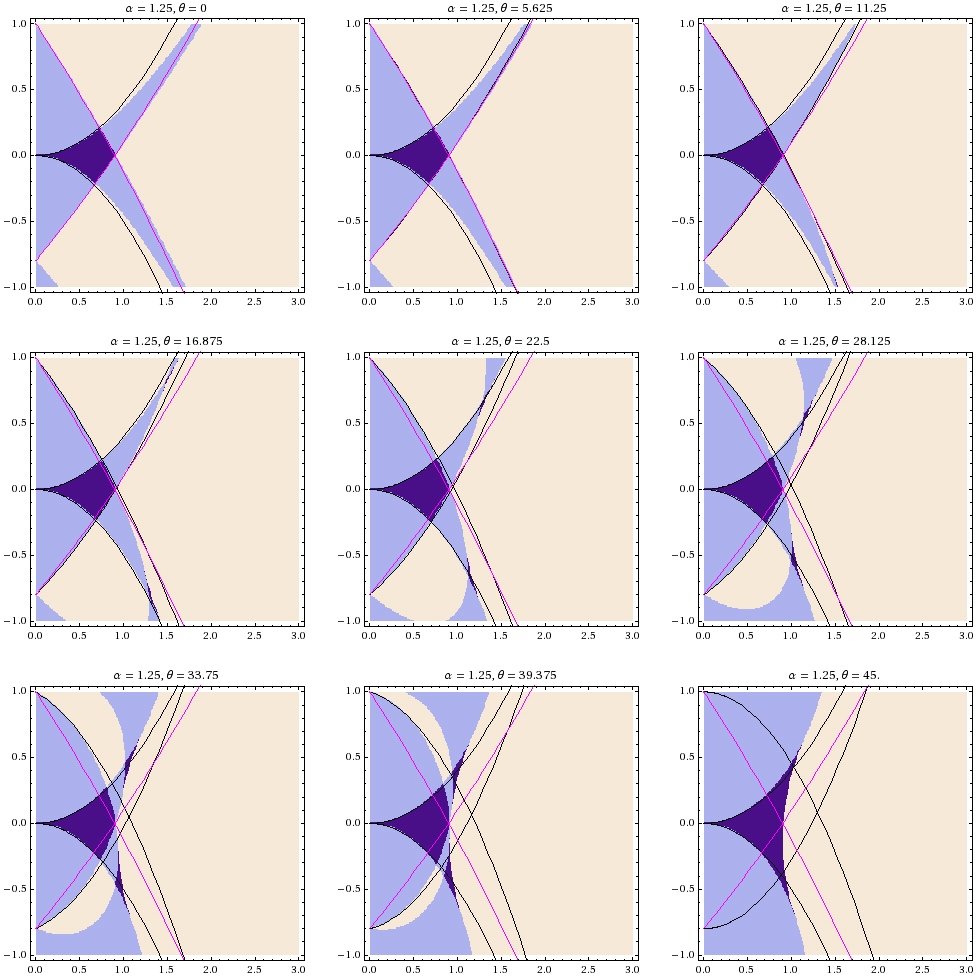}
\caption[]{Stability plots for $\alpha = 1.25$ and $\theta$ between
$0^{\circ}$ and $45^{\circ}$, with a step size of
$5.625^{\circ}$. See the caption for
Figure~\ref{fig:alpha_0.5}.}
\label{fig:alpha_1.25}
\end{figure}

\begin{figure}[H]
\centering
\includegraphics[scale=0.4]{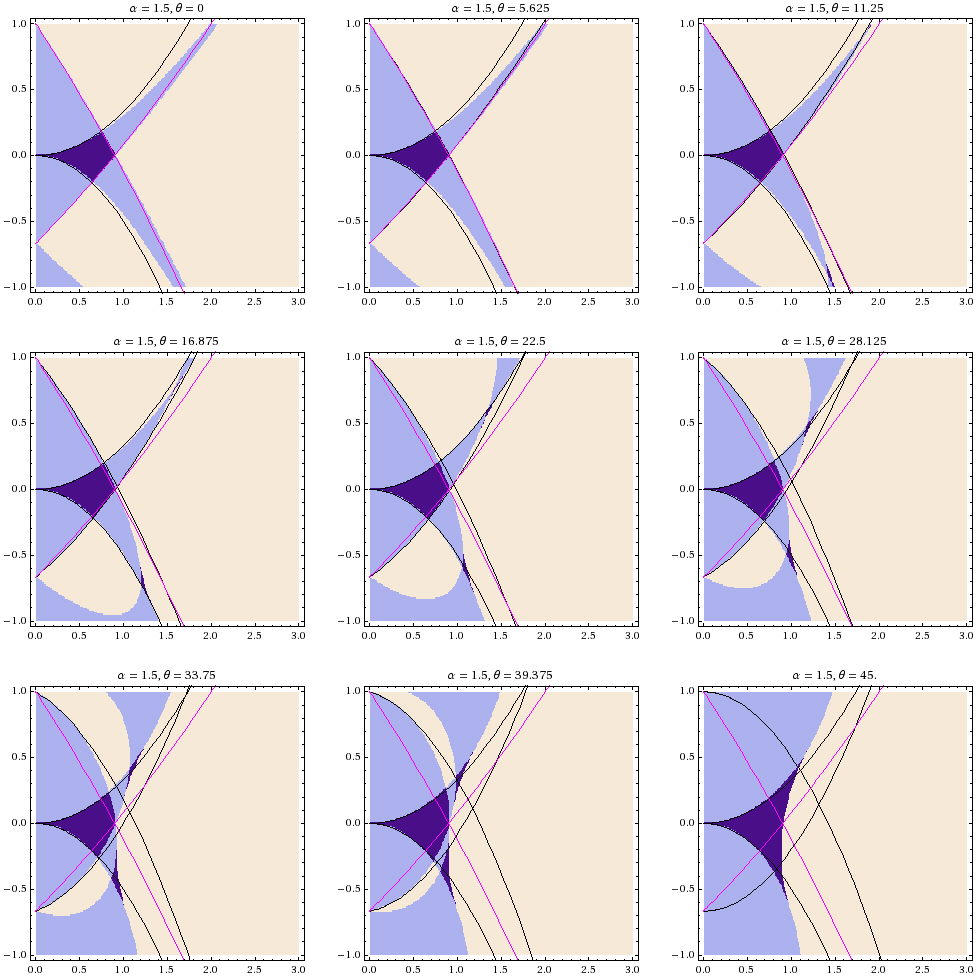}
\caption[]{Stability plots for $\alpha = 1.5$ and $\theta$ between
$0^{\circ}$ and $45^{\circ}$, with a step size of
$5.625^{\circ}$. See the caption for
Figure~\ref{fig:alpha_0.5}.}
\label{fig:alpha_1.5}
\end{figure}

\begin{figure}[H]
\centering
\includegraphics[scale=0.4]{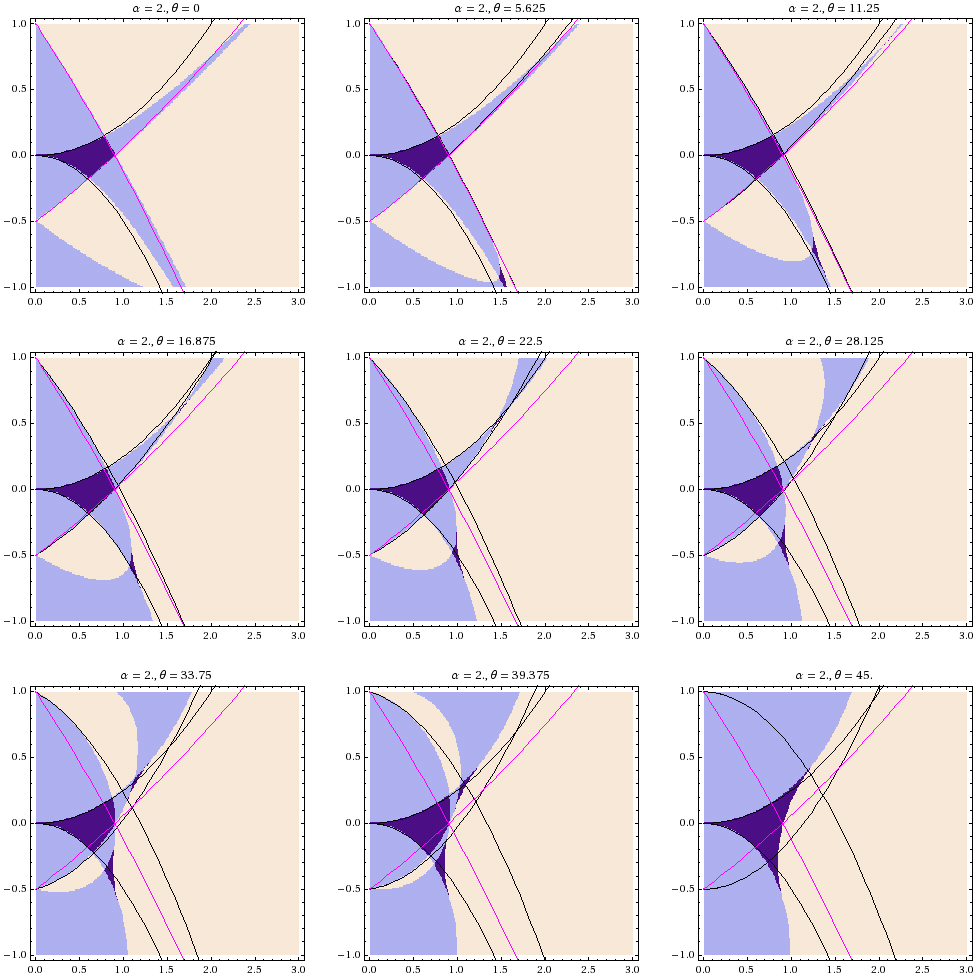}
\caption[]{Stability plots for $\alpha = 2.0$ and $\theta$ between
$0^{\circ}$ and $45^{\circ}$, with a step size of
$5.625^{\circ}$. See the discussion under
Figure~\ref{fig:alpha_0.5}.}
\label{fig:alpha_2}
\end{figure}

\begin{figure}[H]
\centering
\includegraphics[scale=0.4]{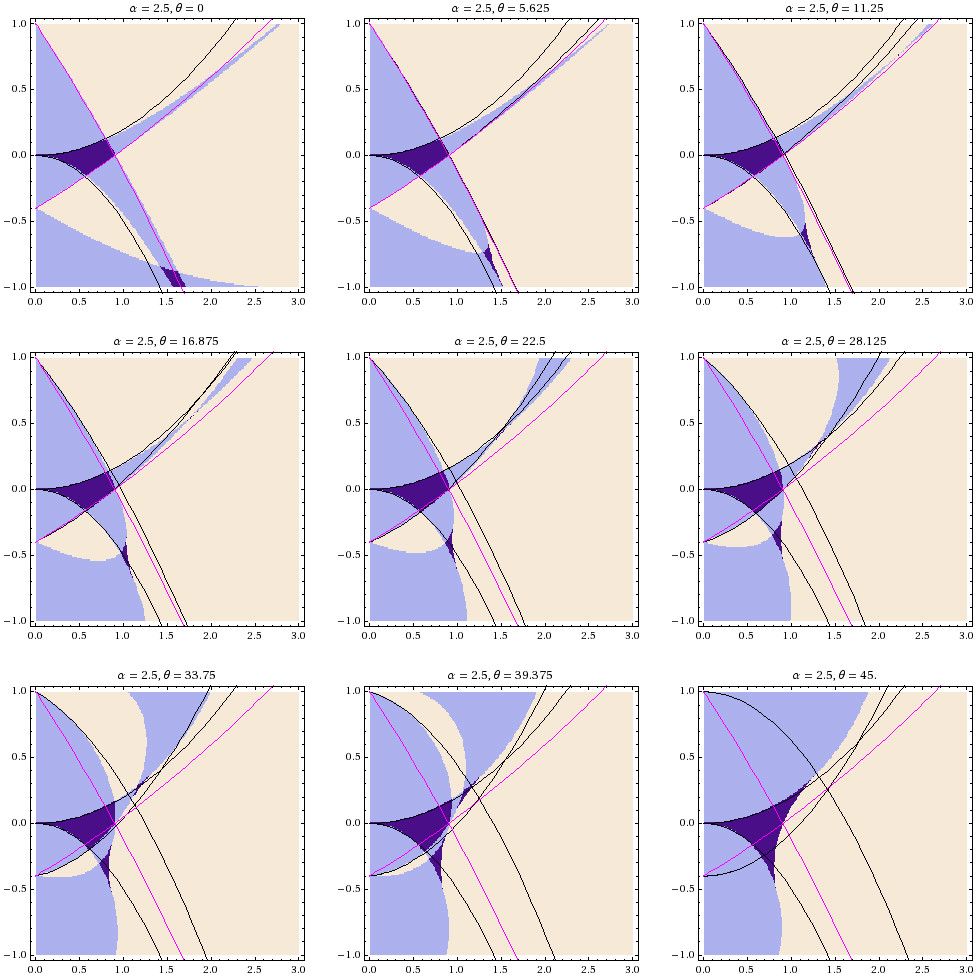}
\caption[]{Stability plots for $\alpha = 2.5$ and $\theta$ between
$0^{\circ}$ and $45^{\circ}$, with a step size of
$5.625^{\circ}$. See the discussion under
Figure~\ref{fig:alpha_0.5}.}
\label{fig:alpha_2.5}
\end{figure}

In Figures \ref{fig:alpha_0.5}-\ref{fig:alpha_2.5},
we present the results of our stability analysis.
We compare the results of the numerical
method of Section~\ref{numstab} (purple areas) to the stability boundaries
obtained by the multi scale perturbation analysis (black lines). We show plots for
a range of angles $\theta$ between the RF and DC axes, and a range of
$\alpha$s (recall equations (\ref{eomfin1})-(\ref{eomfin2})).
A few important conclusions are evident from the plots.
\begin{itemize}
  \item The primary stability region \textit{does not get smaller}
  when a nonzero angle $\theta$ is introduced between the RF and DC
  axes if the other variables are kept fixed.  Such a nonzero angle
  between RF and DC fields represents the case of an asymetric surface
  electrode geometry and/or asymmetric voltages.
  \item Although the primary stability region does not change
  appreciably when $\theta$ is varied, two secondary
  stability regions are
  highly variable, and when $\theta$ has the special value of
  $45^{\circ}$, they join the primary region to result in an exceptionally large
  region of stability.
  \item The curves obtained from multi-scale perturbation theory
  approximate the boundaries of partial
  stability\footnote{If there is a set of initial
  conditions for which the ion motion is bounded and another set
  for which it is unbounded, we call the system \textit{partially stable}. Since
  it is impossible to tune the system perfectly, the
  unbounded solutions of a partially stable operating point of the trap will
  get excited in practice, and the ions will be lost.} quite
  accurately when $q$ is near zero. Regions of partial stability
  become regions of full stability when they ``overlap'' near the primary
  stability region of classical, symmetric traps.
  Unfortunately, in
  this region, $q$ is likely large enough to make the accuracy of
  the approximation unsatisfactory, especially for angles $\theta$ close to the
  critical value of $45^{\circ}$.
  Since, in practice, partial stability means
  instability,
  the small $q$ region, where our multi-scale formulas
  are uniformly accurate, is not of relevance
  to the practical problem of the stability of ion motion.
  \item When we lay the approximate stability boundaries for the
  \textit{symmetric} Paul trap (described by \textit{decoupled} equations)
    on top of the coupled stability plots, we see that these boundaries
    consistently
  \textit{underestimate} the size of the stability region for the coupled system.
\end{itemize}

A relative angle between the RF and DC axes does not change the
boundaries of the primary stability region significantly,
unless the angle is
close to $45^{\circ}$, in which case the stability region is
enhanced. However, the shapes and locations of a pair of secondary stability
regions are highly dependent on the relative angle.

The conclusion of this analysis for the practical problem of asymmetric ion trap design and operation
is as follows: Just as in the case of symmetric Paul traps,
the $q$-$a$ stability plot of the
standard single-variable Mathieu equation is sufficient to allow one
to determine stable operating conditions for
asymmetric surface traps with long RF electrodes.
It is possible to proceed as in the case of symmetric Paul traps, by obtaining a pair of ``$a$''
values for the two radial principal axes of the DC potential,
and a pair of ``$q$'' values for the two radial principal axes of the RF potential.
Ignoring the fact that
there is a nonzero relative angle between the RF and DC axes,
the two decoupled Mathieu equations for these $(q,a)$ pairs
can be used to determine the
stability properties of the coupled system.
This approach does not give precise stability boundaries for asymmetric traps,
but it is ``safe'' in the sense that trap operating conditions deemed stable
by this method will in fact be stable for the coupled system. By ignoring the coupling,
the size of the primary stability region is simply underestimated.

\begin{figure}[H]
\centering
\includegraphics[scale=0.4]{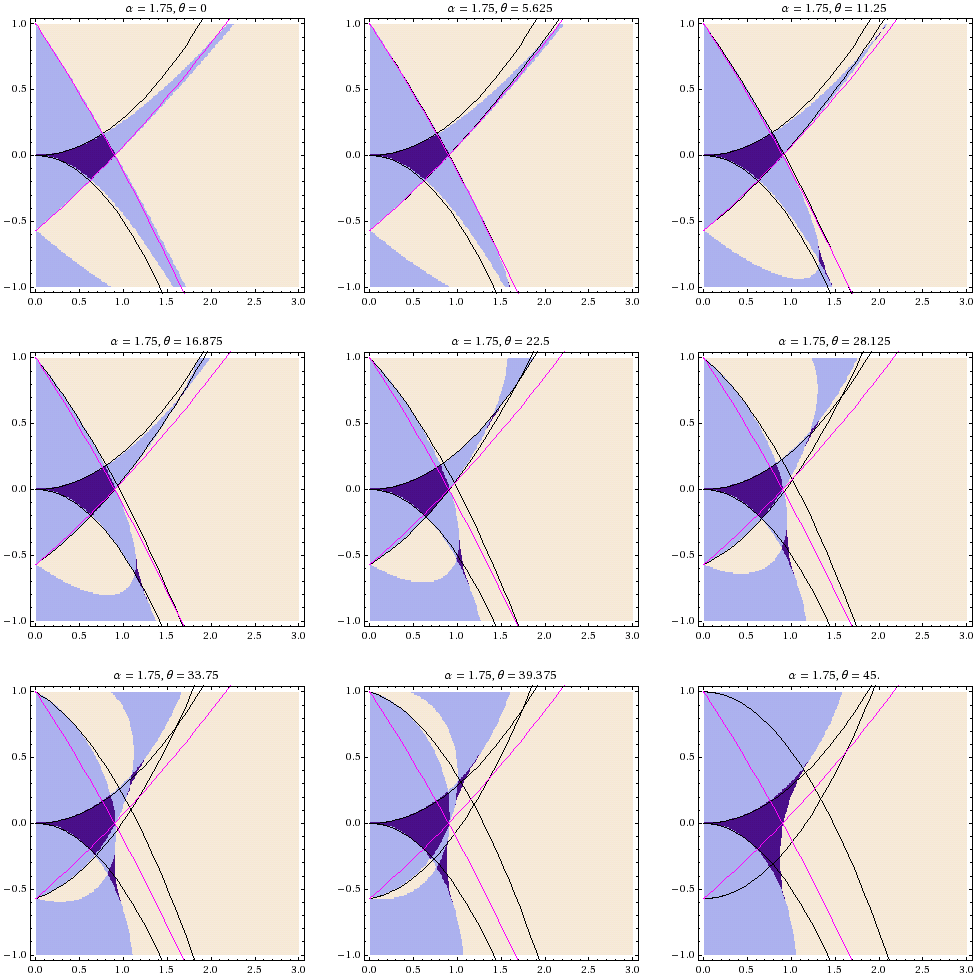}
\caption[]{Stability plots for $\alpha = 1.75$ and $\theta$ between
$0^{\circ}$ and $45^{\circ}$, with a step size of
$5.625^{\circ}$. See the discussion under
Figure~\ref{fig:alpha_0.5}.}
\end{figure}

\paragraph{Acknowledgements.}
We would like to thank Alexa W. Harter, Jason M. Amini, Curtis Volin, and Richart E. Slusher 
for helpful comments and discussions.
This material is based upon work supported by the Office of the Director
of National Intelligence (ODNI), Intelligence Advanced Research Projects Activity
(IARPA) and the Defense Advanced Research Projects Agency (DARPA). All statements of fact, opinion or conclusions contained herein are those of the authors and should not be construed as representing the official views or policies of IARPA, the ODNI, or the U.S. Government. US Army Research Office contract support through W911NF081-0315, W911NF081-0515, and W911NF071-0576 is acknowledged.

\bibliographystyle{unsrt}	
\bibliography{myrefs}	

\end{document}